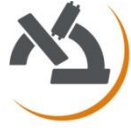

Faculty of Engineering

Department of Computer Engineering

# A Simulation System for Energy Harvesting Networked Tags (EnHANTs)

Sapir Erlich

Alexander Lavzin

Adrian Segall

Michal Ireen Yomtovian

# Contents

# Abstract

This project is about EnHANTs networks [1],[2]. These networks have self-sustainable energy devices and the communication is performed over wireless links between neighboring devices. Since the devices are very limited in terms of energy, sending the data in these networks has to be as effective as possible and must take into consideration the available amount of energy. Therefore, it is necessary to use algorithms that are specifically designed for this type of networks.

In this book we propose strategies for a specific case of these networks. The main part of this project is a MATLAB simulation built in order to examine the behavior of such networks. In the last section of this book, we describe a set of runs of different topologies, in order to test the influence of certain strategy parameters on the behavior of the system.

# Acknowledgments

First, we would like to thank Professor Adrian Segall for his guidance, from the beginning of the project and throughout the year. Prof. Segall and Mr. Alexander Lavzin have helped us and have answered all questions we have had. Without doubt, we have expanded our knowledge about sensor networks. We have had the privilege to work on an interesting subject and to take part in furthering the research they have started.

In addition, we would like to thank our families, especially our parents, who have supported us during the last year and all through the four years until graduation. Thank you for the help and for the huge support on a personal and academic level. We very much owe you our achievements!

<div style="text-align:right">Michal Yomtovian and Sapir Erlich</div>

On a personal note, I would like to dedicate this book to my father – Yossef Yomtovian – who passed away a few days after the end of my last academic year.

<div style="text-align:right">Michal Yomtovian</div>

# The Problem

## Sensor Networks

Our work deals with EnHANTs sensor networks [1],[2]. Every EnHANTs system includes:

1. Data producing sensors: the system source nodes, which produce data packets at some rate.
2. A destination node.
3. Intermediate nodes that are part of paths for transferring the data packets from the sources to the destination.

To each source we designate two possible paths for transmitting a data packet. The time is divided into slots and during each slot each source transmits packets via one of these paths. The communication is wireless between neighboring nodes. Neighboring nodes are pairs of nodes that can hear each other directly.

In this system, the intermediate nodes have limited batteries and they can harvest energy into the battery. Each node harvests energy at a certain rate (which can vary during the running time) and every transmission, receiving or hearing a packet, consume a certain amount of energy. If the energy level in the battery of a node is not sufficient to execute the required action, the action is not executed and the packet is dropped.

The destination node is assumed to be energy unlimited and to have a powerful transmitter that can be heard by all network and source nodes. It also has the ability to control the activity of the system by deciding which routing paths will be active in each slot. In addition, it dictates changes in the data producing rates at the sources. The control strategy defines the algorithm employed by the destination to decide upon the input rates and upon the active routing paths.

# Goal

The final goal is to find a good strategy, so that the network throughput (the amount of packets that arrive at the destination) will be as high as possible, with as few losses as possible. Our MATLAB program simulates the behavior of the system with one of several strategies. We then run several scenarios with several sets of parameters to test the behavior of the system in terms of throughput and lost packets.

# Theoretical background

*This part is based on the HDR article [3], on notes about it -written by Alexander Lavzin and on several unpublished related notes [4].*

EnHANTs- Energy Harvesting Active Networked Tags - are networks with self-sustainable energy harvesting devices that communicate with neighboring devices over wireless links. In such networks, node energy increases via harvesting and is spent by data and control transmission and reception.

## Basic model
Consider a diamond network with 4 nodes.

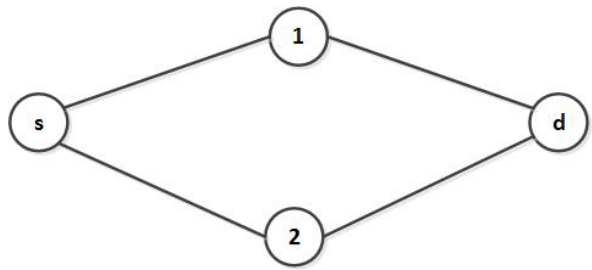

Node *s* is the source node and *d* is the destination. Node *s* generates data to be transferred via the energy-harvesting nodes 1 and 2 to the destination *d*. Data is sent in data packets of fixed size. Nodes 1 and 2 have a limited amount of energy. Energy is regained with energy-harvesting rates, which may vary throughout the operation time. The source and the destination are assumed to be energy unlimited. The time is divided in slots. The activity of the network nodes is defined before the start of each slot, and throughout the slot it cannot be changed. Thus, any change in the activity of the source or other nodes is possible only at the end of a slot. Every wireless transmission by the source is overheard by both forwarding nodes, but in any given slot, only one of them, referred to as the *active* node, is forwarding the packets to the destination. Only the *active* node spends energy by receiving the packet and forwarding it. The other node is said to be *inactive*. As the routes are one node long, if node 1 or 2 is active, we also say that we use route 1 or 2 respectively.

At the end of each slot, the intermediate nodes inform the destination node of their current energy levels. Based on this information, the destination node decides what route to use and informs the forwarding nodes accordingly. The decision is sent to the node by the destination node just after it receives the status messages, in a control message that is referred to as a switch-command message. We assume that it takes negligible time to send the status messages, to receive the switch command and to apply it at the forwarding nodes. If a packet is dropped for any reason, the information it carries is lost. The nodes do not store packets in order to retransmit them if the delivery fails. We use the following notations:

The period between two consecutive switches in the same direction is referred to as a *cycle*.

$e_u$ - energy harvested by node *u* in each slot (mJ)

g - number of data packets generated by the source node in each slot

$c_t$ - energy to transmit a status packet (mJ)

$c_r$ - energy to receive a switch command packet (mJ)

$D_r$ = energy to receive a data packet (mJ)

$D_t$ = energy to send a data packet (mJ)

$D = D_r + D_t$. energy to receive and transmit a data packet (mJ)

$B_u^-(i)$ - battery level at node u just before the end of slot i in units of energy (mJ)

$B_u(i)$ - battery level at node u at the end of slot i, in units of energy (mJ)

$B_{max}$ - maximum battery level (mJ)

$B_{min} = c_t + c_r$ - minimum battery level that allows transmission of control messages (mJ)

$h_1$ - threshold to switch from route 1 to route 2 (mJ)

$h_2$ - threshold to switch from route 2 to route 1 (mJ)

$1_z$ = 1 if z is true and 0 otherwise

We would like to balance the transmission between the two nodes so that the amount of packets arriving at *d* will be maximal.

# HDR logic

The proposed algorithm, referred to as the Hysteresis-Driven Routing (HDR) Algorithm [3] is as follows. The destination assigns as active the node with the higher energy level, but in order to avoid fast oscillations causing a high control overhead, the routes switch is performed only when the energy level at the inactive node exceeds the level at the active node by a certain threshold. The threshold for switching in one direction may be different from the one in the opposite direction. Since activity can change only at the end of a slot, if the threshold is reached during a slot, the actual switch time is at the end of that slot.

Energy Parameters definitions:

$H_{Dr}$ = energy required to receive the header of a data packet and identify if the packet is to be received and processed, included in $D_r$.

$H_{Cr}$ = energy required to receive the header of a control packet and identify if the packet is to be received and processed, included in $C_r$.

$A_r$ = energy required in Idle to assess whether a control packet is being sent, included in HCr.

As we continue to discuss more complex systems, with several sources and longer paths, we need to consider some new issues:

**Packet overhearing**

All of the communication are wireless. Therefore when a node transmits a data packet, it is overheard by its neighbors. There are 3 types of possible neighbors of a node on the active path:

1. Next-Hop - The next node on the active path from source to sink.
2. Previous-Hop - The previous node on the active path from source to sink.
3. Cross-Source - A node which is on an active path of another source.

*Note* : we assume that the paths are neighbor-loop free, namely that no node on each path has a neighbor, other than its next-hop and previous- hop, that is on the same path.

When a Next-Hop neighbor overhears the transmission, it realizes that the packet is intended to itself, and continues to receive the entire packet. However, if a node of any of the other types overhears the transmission, it analyzes the header and stops receiving the transmission. This process costs the node $H_{Dr}$ . A link between two nodes that is not Next-Hop to either of them is referred to as a NonPath link. The destination node has two main control missions for each source: switching from one path to the other and adjusting the data rate.

# Feedback

**Drift**:

The node with minimal energy on a given path will be referred to as the *critical node* of that path. Recall that for a given source, the number of slots between two consecutive switches in the same direction is referred to as a *cycle*. For a given source, the drift in energy levels in a given cycle is measured as the difference between the energy at the critical node on the active path at the end of the cycle and the energy at the critical node on the active path at the beginning of the cycle. The drift divided by the number of slots in the cycle is the slope of the energy level of the active path during the cycle.

**Feedback logic**

In principle, if the energy slope is upwards (positive slope), the input should be increased, and if the slope is downwards (negative slope), it should be decreased. Increase/Decrease of the input rate will be referred to as the Input Shift. In our strategies small energy drifts are tolerated. One approach, used in the *FeedbackWithDangerZones* strategy, is to arbitrarily select a value for the minimum slope that causes an Input Shift decision. In our algorithm, this variable is referred as *Allowed Drift* and its default value has been empirically selected as 0.003 mJ/slot. A hopefully better approach, employed in the *Estimated Slope* algorithm described below, is for the destination to estimate the effect that an Input Shift in the next cycle would have on the behavior of the system and employ this estimate for the decision if such a Shift is warranted.

**Estimated Slope**

We define the drift for a given cycle as the difference between the sum of the harvested energy and the sum of the spent energy in the critical nodes on the two paths. The total energy drift for a given source is the difference between the harvested energy and the spent energy. As such, we shall use the following estimate of the change in drift of the total energy due to a negative Input Shift of one packet per slot, calculated by the sink at the end of a cycle:

$$\delta = D * I + (I_1 * 1_{CN1} + I_2 * 1_{CN2}) * HD_r$$

If we decrease the transmission rate on one packet, the energy consumption per slot of the critical node in the active path would decrease by D. Therefore, the total energy consumption used for data on the whole cycle would decrease in D*I. The other term is the total inline overhearing energy saved by the critical nodes due to the decrease of the source rate. The energy saved by the critical node on Path 1 is $HD_r$ times the number of slots Path1 is active, provided the critical node is not the last on Path 1 (the last node on the path does not overhear other transmissions of the packet, because a node on the active path overhears the packet only when the next node on the path transmits a packet). Similarly for the other Path. The critical nodes may change during the next cycle, but the sink has no way to know this a priori, so the best guess is to use the current critical nodes. The number of slots in a cycle is proportional to the input rate, but it also depends on the harvesting rates, which are unknown to the sink. Therefore the sink needs to employ its best guess, namely the previous cycle length. The natural way to use this estimate is to dictate an Input Shift, if, according to the estimate, it will bring the drift closer to zero than if there is no Shift. Denoting by *s* the average drift per slot per node calculated at the end of an interval, the total drift in the previous cycle is 2*I*s. Therefore this amounts to

$$\text{Shift} = \begin{cases} 1 & if\ 2*I*s > \delta/2 \\ -1 & if\ 2*I*s < -\delta/2 \\ 0 & otherwise \end{cases} \quad or \quad \text{Shift} = \begin{cases} 1 & if\ I*s > \delta/4 \\ -1 & if\ I*s < -\delta/4 \\ 0 & otherwise \end{cases}$$

Therefore, the value of delta/4 can be used as the Allowed drift of the previous feedback strategy.

**Danger Zones**

Input shifts can be also used for keeping the critical node's battery level away of the boundaries, regardless of the current drift. Thus, this type of input shifts can be performed also when the destination node does not decide on having a path switch. There are three types of danger zones:

- $B_{max}$ zone- when the critical node is close to $B_{max}$ (the energy level is above $0.9*B_{max}$) indicating that a higher transmission rate can be used. A positive input shift should be made.
- High danger zone- the critical node is very close to $B_{min}$ (the energy level is below $B_{min}+D*High\_Danger$). In order to avoid packet loss, a negative input shift should be made.
- Low danger zone- the critical node is close to $B_{min}$ (the energy level is above the high danger zone, but below $B_{min}+D*Low\_Danger$). In order to avoid going into the high danger zone, a positive input shift should be prevented.

The High_Danger and Low_Danger are parameters of the strategy, empirically selected as 90 and 200 respectively.

# Strategies

In order to examine the suggested solutions, we have constructed a simulation system on Matlab. We combine the HDR and Feedback logics previously described, to define the 3 strategies included in the simulation:

1. **No Feedback**– At the end of each slot, the destination node decides which one of the two paths of each source will be the active path at the next slot.

   No feedback logic is included, so the sending rate of each source is a set parameter at the beginning of the simulation and it does not change until the end.

2. **Feedback with Danger Zones** – At the end of a slot, in addition to the decision about the active path, the destination node decides for each source whether it needs to raise or reduce its transmission rate, depending on the paths' energy level. A transmission rate change is decided if the battery level is in a "danger zone" or if the energy level drift of the source reaches some threshold value that was determined at the beginning (the Allowed Drift parameter).

3. **Estimate** – similar to the "Feedback with Danger Zones" strategy, but the drift threshold used to change the transmission rate of a source, is estimated each time, depending on the current energy level of the routing paths of the source.

# Assumptions for Simulation

- The nodes are static.

- The destination node learns the energy levels of all nodes in zero time at the end of each slot.

- The destination node has a powerful transmitter that can send control messages to all nodes in zero time.

- The receiving and transmission rates of the intermediate nodes are sufficiently large such that whatever is sent to them is immediately sent out. No collisions upon arrival and no accumulation of packets occur.

- Packets arrive at random times within a slot. If there is enough energy, the received packet is transferred, if not, it is discarded.

- The energy level should never go below $B_{min}$ before the end of a slot (after all data packet are sent, before sending status packets), so there is always enough energy to send status packet and to receive command packet at end of slot.

# Simulation Block Diagram

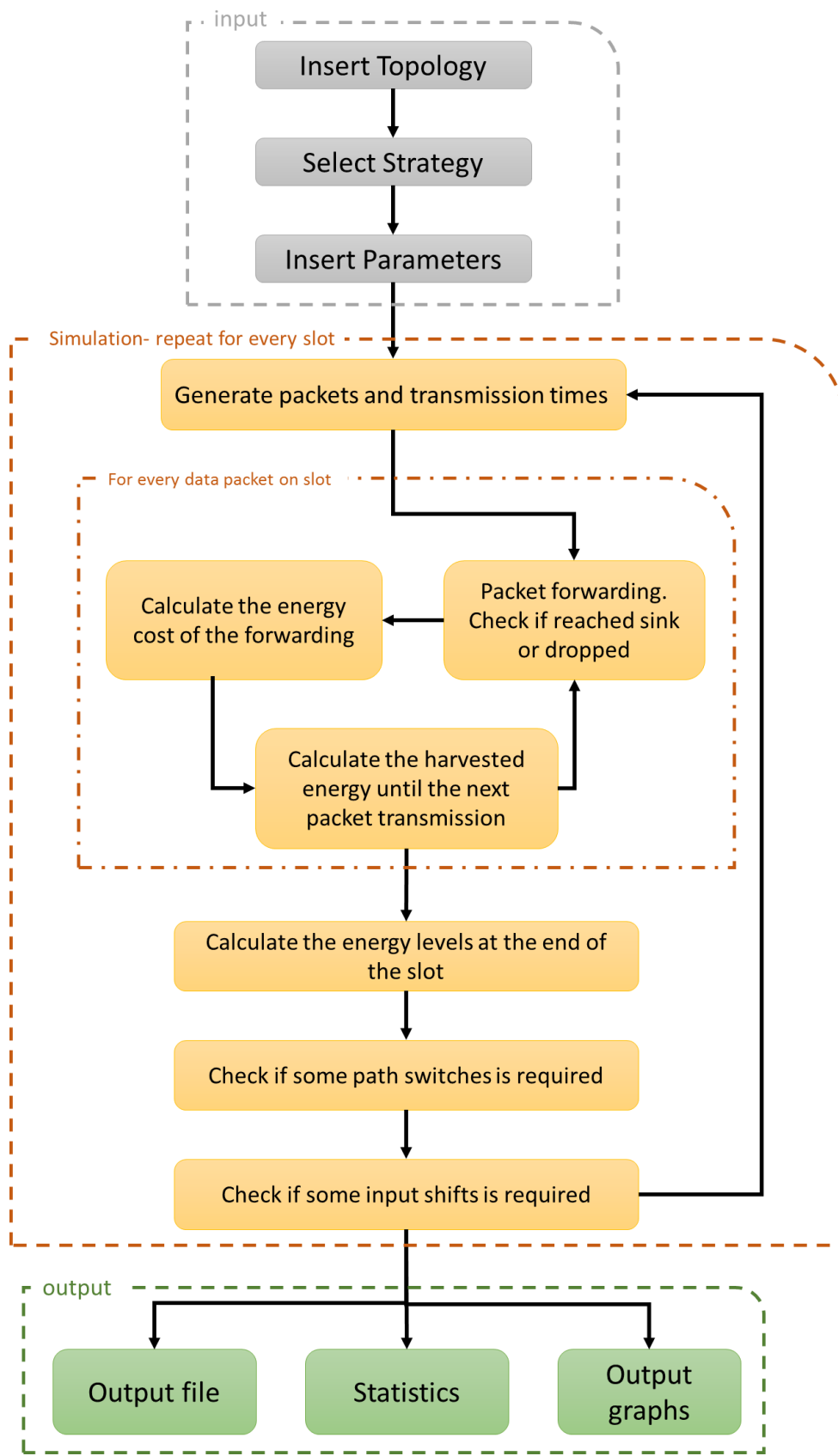

# Simulation Description

**Debug options:**

Exact Same Run for Simulation.m:

Run with the same parameters and topology as the last run, packet schedule is taken from the last output file, so the run is exactly as last run.

Same Parameters as Last Run:

Skip all the other input screens. Run with the same parameters and topology as the last run. The rest of the simulation is a normal run [packet schedule is made in random for each slot (as in a normal run)]

Print switch times:

Print the system state (slot number, new active paths, transmission rates, harvesting rates, battery levels) on the first slot and on every slot the destination decided a switch.

Strategy debug:

Used for feedback strategies. Each time the strategy decides a switch-time input shift should happen, the simulation prints:

- The slot number.
- The number of the source having the shift.
- The shift type (back/$B_{min}$/$B_{max}$) and direction.
- The number of sources that use current min-node (if the shift is a back shift).

- The number of sources that use previous min-node (if the shift is a back shift).
- The drift (if the shift is a back shift).
- The battery level of the min-node.

For back shift on Estimate strategy, the calculated drift threshold (delta/4) is printed.

It also prints a message when a +1 back shift is reduced to 0 by Low Danger Zone. When the destination decides to make the shift (the shift is not prevented by a shift pause), print the slot number, the shifted source number, the direction of the shift and the new transmission rate.

<u>Non-switch shift debug</u>:

Used for feedback strategies. When the destination decides a non-switch shift, print the slot number, the shifted source number, the direction of the shift and the new transmission rate.

# Input

**Topology:**

The user enters the number of sources and the number of intermediate nodes. A topology building screen is then shown, including all the nodes: Source nodes (in black) and intermediate nodes (blue), where the number of each type matches the values the user entered, and one destination node (green). In addition, a window appears with the following topology define options:

<u>Topology as before:</u> if a saved topology with the same number of sources and intermediate nodes exist, this topology is loaded.

From scratch:  the user enters the route (numbers of the intermediate nodes forming the path) of each path for every source in the topology. After entering each path, it will be shown on the topology graph.  The user is then asked if there are more links to add.  If the user chooses yes- the user adds a link by typing the two node numbers that the link connects.

Move node: Used for changing the location of the nodes in the topology graph.
The user enter the number of the node to move, and then chooses the new location on the graph window, using the mouse pointer.

Change source path: the user chooses the source to change, its path to change (1 or 2) and then enter the new route of the path.

Add more links: for adding the network links that don't take part in any routing path.

The user adds the link by entering the numbers of the two nodes the link should be connecting.

Remove links: for removing non-path links. The user removes the link by entering the numbers of the two nodes the link is connecting.

After each action is done, the topology is updated and shown, and the user can choose another option, or choose "Finished" to finish.

**Strategy:**

The user then chooses a strategy for the run, the options are- "No Feedback", "Feedback with Danger Zones" and "Estimate".

**Parameters:**

A set of input dialog boxes is shown, for the user to enter the required parameters for the simulation.  The parameters are divided to four sections, and before each section, the user can choose whether to use this section's parameters from the last run (if exist), or to enter new values.

The parameters are:

Energy-Cost Parameters:

*Send_Bit_Energy*- Energy required to transmit a single bit [uJ]

*Receive_Bit_Energy*- Energy required to receive a single bit [uJ]

*Data_Packet*- Length of a control packet [bit]

*Control_Packet*- Length of a control packet [bit]

*Data_Header*- Length of a data packet header [bit]

*Control_Header*- Length of a control packet header [bit]

*Equiv_Detection*- Equivalent in Bits for signal detection (To calculate the energy "cost" of a detection if a control packet is sent)

Simulation Parameters:

*No. of slots*- Number of slots to simulate.

*No.of SS Slots*- Number of slots collecting steady state information.

*High_Danger*- Zone of high danger = D * High_Danger. Battery is very close to Bmin.

*Low_Danger*- Zone of low danger = D * Low_Danger. Battery is close to $B_{min}$.

*Max_Shift_Pause_Up*- How many switches is the pause in shifts-up after performing a shift.

*Max_Shift_Pause_Down*- How many switches is the pause in shifts-down after performing a shift.

*Max_Shift_Pause_Slot*- How many switches is the pause in shifts after performing a shift.

*Allowed Drift*- Used for the "Feedback with DangerZones" strategy. The drift threshold for input shifts.

Source Parameters:

*Initial Source Rate (g) for source i*- The transmission rate of source i in slot 1 of the simulation

*Threshold while path 1 active for source i*-The threshold for switching the routing of source i from path 1 to path 2.

*Threshold while path 2 active for source i*-The threshold for switching the routing of source i from path 2 to path 1.

*Initial route for source i*- The active path of source i in slot 1.

Harvest-Node Parameters:

*Harvesting rate (e) for node i*- The energy harvesting rate of node i. if the value is greater than 100, the rate is not constant, and read from a specific file.

*Initial battery value for node i*- The battery level of node i at the beginning of slot 1.

*Maximal battery value for node i*- The maximal energy level possible for node i.

# Simulation

Before simulating the first slot, a calculation is made to "verify" the initial route, to make sure that no path switches are required at the beginning of the first slot. The rest of the simulation is done by performing the following parts for each slot:

"Listening nodes": mark all the active nodes on this slot- the nodes on the active paths- those are the nodes that will use energy on this slot (before the status messages at the end of the slot). The non-active nodes don't send any data packets in this slot, so they will not use energy (before the status messages) even if a message is sent from a neighbor node.

**Generate packets**: the total amount of packet to send in this slot is calculated as the sum of the transmission rates (g) of all the sources. A random value from 0 to 1 is assigned to each packet, indicating its transmission time at the slot (=when the packet is sent). The packets are sent in the order of the generated time values. Since it is assumed that it takes zero time for a single packet to arrive at the destination, there will not be any collisions between packets, so the simulation can simulate every packet forwarding separately.

The **packet forwarding** is calculated for every packet separately, by the packet sending order. The following calculation is done:

- Decrease $H_{Dr}$ from the energy level of every "listening" node that is a neighbor of the source that transmit the current packet, and has enough energy, except for the first node on the active path. If a node is a neighbor of the source, but it is not on the source's active path and it has enough battery, it overhears a message that is not intended for it and loses $H_{Dr}$ energy.

- Then, for every node on the active path:
    - Try to transmit the data packet to the node. If the node energy level is above $D + B_{min}$, the packet arrives at the node- decrease D from its energy level.   (If not- the transmission failed- the packet drops.  Break loop and move to the next packet)
    - Then for every "listening" neighbor node of the current node, that has enough energy, except for the next node on the path (but include the previous node on the path) - decrease $H_{Dr}$ from the energy level, because of overhearing.
- If the packet has successfully passed all the path nodes and arrived at the destination node, add it to the throughput count.

After sending all of packets generated in the slot, the intermediate nodes report their energy level to the destination through a status message, which cause an energy decrease of Ct in all the nodes. If the energy level of a node is less than Ct, it can't send the status message, and the sources that have this node on their paths is affected.

The destination node might send a control message, so all nodes should have sufficient available energy to receive the control packet. If a node does not have enough energy, the simulation will stop, because the algorithm assumes that all nodes follow the instructions from the destination.

The following logic of switches and shifts is done separately for each source:

**Path Switch**:

For each path (both the active and the inactive paths of the source) the node that has the minimum battery level in that path is found, using the Min_On_Path function, which omits nodes that appear on both paths from being included in the minimum-node calculations. If the energy difference between the minimal node on the inactive path and the minimal node on the active path is higher than the corresponding threshold, a path switch is decided for the source. If the strategy is "No Feedback", skip the following parts that refer to input shift, which is not relevant to this strategy.

**Input Shifts**:

If, according to the previous part, a switch of the paths of the source will not be made on this slot, we check if a <u>non-switch shift</u> is required:

- Assume shift up is required- but cancel the shift up if any battery on the Active Path does not increase or stay at same level, or if the path energy level is not in the "$B_{max}$ zone" (which is above $0.9 * B_{max}$).
- If no shift up is required, maybe a shift down is needed. Usually shift down is done on a switch-shift, but if the energetic bottle-neck is a node which is on both paths of the source (Omit), then it is ignored by the shift-switch logic, but treated here.

If a switch will be made at the source on this slot, we check if a shift is also needed:

- First, according to the chosen strategy, find out if an Input shift is needed:
    - If the energy level is in the high danger zone (energy is below $B_{min}$ + High_Danger * D ), a -1 shift is needed.
    - If the energy level is in the low danger zone (above high danger, but below $B_{min}$ + Low_Danger * D), a +1 shift will be prevented.
    - If the minimal battery of the inactive path is in $B_{max}$ zone, a +1 shift is needed.
    - If a no danger zone shift is needed, try for a back shift:
        - Calculate the drift

- In "estimate" strategy- evaluate delta, set Allowed_Drift as delta/4.
- In "feedback with danger zones" the Allowed_Drift is a set parameter.
- If abs(drift)>Allowed_Drift, a back shift is needed, the direction of the shift is sign(drift).

- If shift is needed but the matching Shift Pause is not zero, it prevents the shift from occurring.

If the destination sends a control message due to a switch or a shift on any source, the energy level of all nodes is decreased by Cr.

If there is a node that can't receive the control packet because its energy level is too low, the simulation stops.

# Output

**Output graphs**

There are three graphs shown at the end of the simulation:

Topology graph

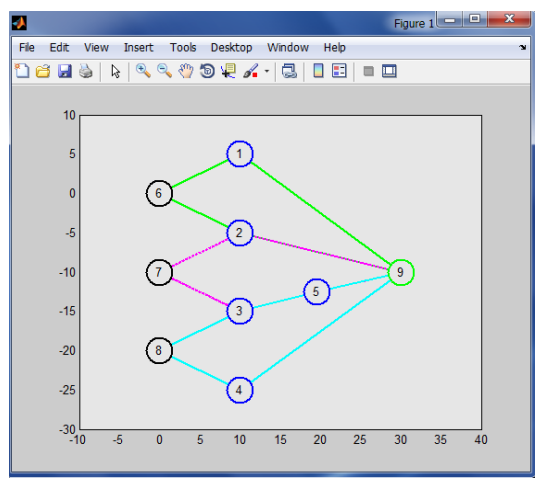

Energy level graph

The graph shows the energy level of each intermediate node throughout the run.

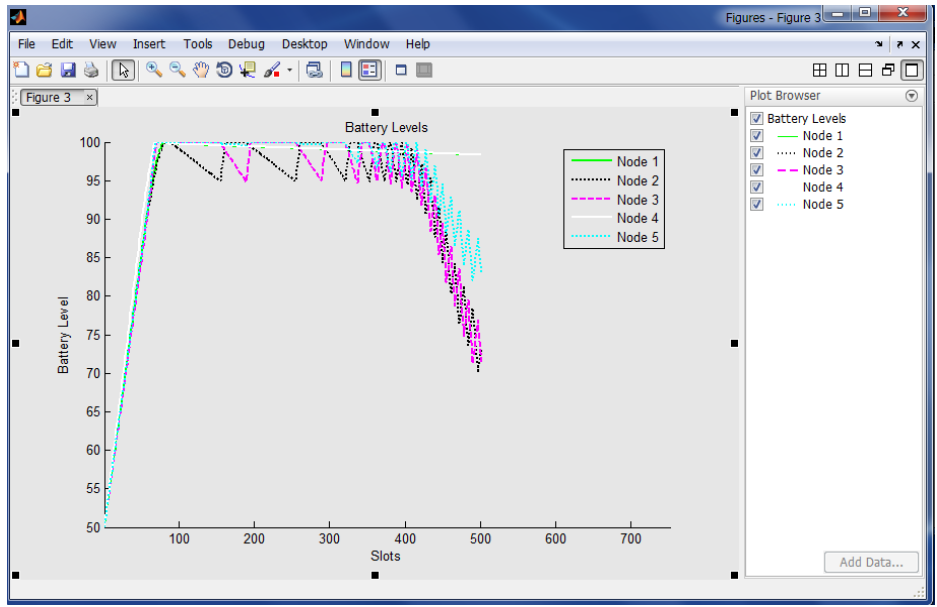

Transmission rate graph

The graph shows the transmission rate of each source throughout the run.

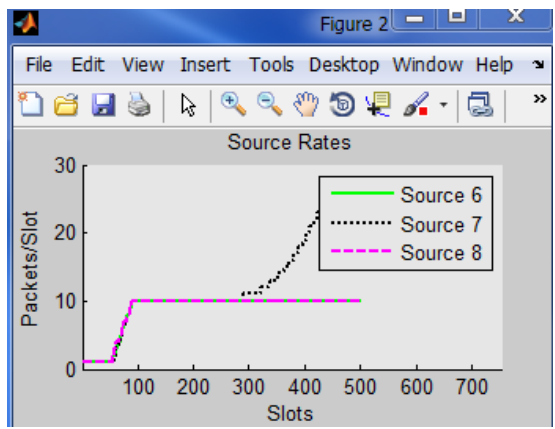

**Statistics**

Some statistics about the simulation are collected along the run, and displayed in the command window at the end of the run, right after the simulation parameters are printed:

- Number of packets sent from each source
- Number of packets received at the destination from each source
- Number of lost packets from each source
- Number of path switches at each source
- Routing state statistics-The percentage of time each routing state was active during the run. A routing state is denoted by the numbers of the active paths on that state. (For example [1, 2, 1] is a routing state of a topology with three sources.)

**Output file**

An output file is created during the run of the simulation. This file contains the packets schedule in each slot, the energy level of each node in the end of each slot, all the debug messages, the parameters of the run and the statistics that were collected.

# Correctness

First, we have examined the basic "diamond topology", as described in [3]. This topology includes one source and two intermediate nodes, each node creates a different routing path. We have run the simulation with the "No Feedback" strategy, which was the strategy presented in the article, and compared the simulation result with the theoretical analysis of the performance.

Next we have run the simulation on a large number of topologies, scenarios and parameters :

- We tried it on basic topologies, where we could try to evaluate the proper transmission rate when using "No Feedback" strategy, and could check if the results we got match those values.
- We compared the results on some symmetric cases of topologies, to see if the results are also symmetric, and compared them with the results for other similar but asymmetric topologies, to see what differences it cause.
- We have tested the simulation system on networks with extreme (low or high) parameters, and checked if the system behaves as expected.

The simulation system has been debugged and improved throughout those tests, so the current simulation system is much more reliable.

# Parameter Examination Runs

We have investigated how different parameters values influence the simulation output on four different topologies. The outputs consist of the total throughput, the number of path switches, fairness (whether the same amount of packets is sent from each source to the destination in symmetric systems) and the energy levels graphs.[1]  Both strategies- "Feedback with Danger Zones" and "Estimate" were investigated.

## The Examined Parameters

- **Initial Routing**- Routing is the set of the active paths (one active path for each source) in a given slot. Every source has a parameter called "Initial Route" that defines which of its paths is active in the first slot of the simulation. The set of the initial routes of a simulation run is the "initial routing" of the run. Every possible initial routing of the topology was checked, and the initial routing which got the best throughput is selected as the initial routing for the following parameters runs of that topology.
- **Shift Pause-** Three parameters [max_shift_pause_up, max_shift_pause_down, max_shift_pause_slot] that determine the minimum time period (in slots) between two successive shifts. On each run, all three were given the same tested value. The tested values are: 1,2,3,5,8,15.
- **Thresholds**- The threshold value indicates when the source should switch its active path to the non-active path. Each source has two

---

[1] All simulation results we discuss here, appear in the appendix of this book.

thresholds- one for switching from the 1st path to the 2nd path, and one for switching from the 2nd path to the 1st path. In each run, all thresholds were set to have the same tested value. The tested values range from 3 to 5.

- In the part where the examined strategy is "Feedback with Danger Zones", we check the parameter "**Allowed Drift**" as well, this parameter determines the minimal drift size causing a shift in this strategy.

  This parameter is tested with the values- 0.3, 0.03, and 0.003.

## Remarks and Definitions

For every parameter examined, the main quality factor is the total throughput- the number of data packets that arrived at the destination node. The result of the simulation may vary because of the randomness in the transmission time of the packets at each slot, so for every tested parameter value we ran the simulation twice, and used the average of those as the result.

In some topologies, in the beginning of the simulation there is a stage where the battery level of the intermediate nodes is around $B_{max}$, because the transmission rate is not sufficiently high. Right after this stage, the control occurs and increases or decreases the transmission rate according to the nodes battery level.

Each run was 10000 slots long.

# Topology #1

The basic topology is constructed out of a single source and two intermediate nodes- each of the two is part of a different routing path.

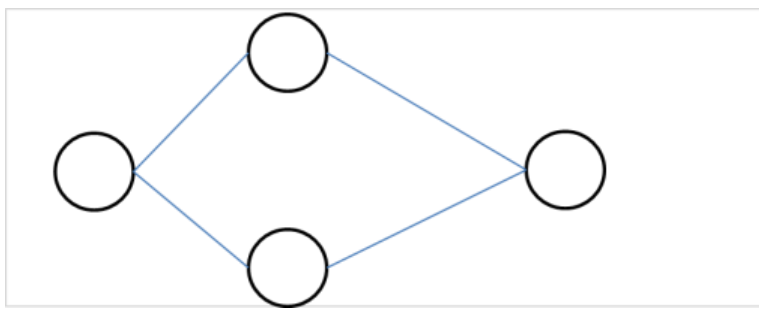

## **Estimate:**

**Initial Routing-** The two initial routing options give symmetric results, which make sense, because the topology is symmetric. Despite the Symmetry, the routing state statistics was not perfectly balanced between the two routing states (in all runs, one state was a bit more common than the other), depending on the initial routing. The initial routing chosen for the following runs was routing through the upper path (though there is no clear preference for either).

**Shift Pause-** As the parameters' values gets higher, less packets are sent, less switches occur, and it takes more time until the control take place.

Shift Pause =1

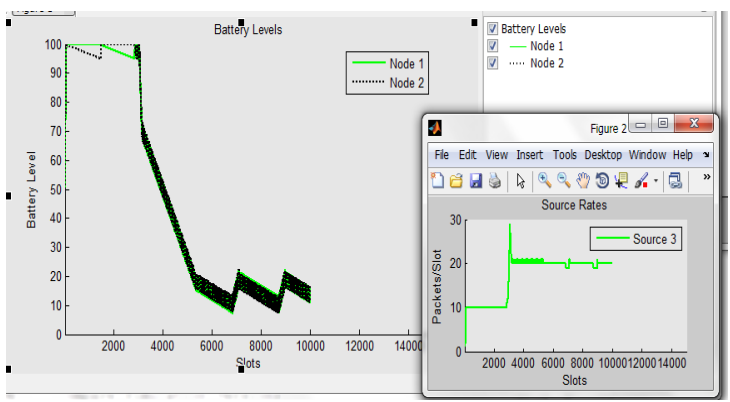

Shift Pause =5

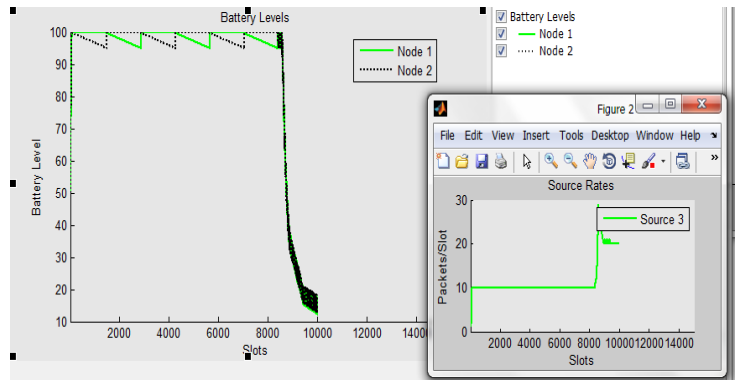

For the highest values tested (8,15) the control didn't take place at all.

Shift Pause =15

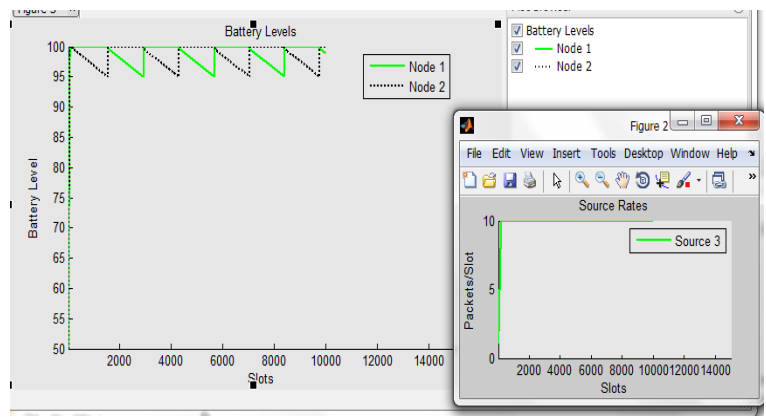

**Threshold-** As the parameter value gets lower, more switches occurs, more packets are sent and the control takes place earlier.  The energy level graphs get less "wavy" as the parameter gets higher.

Threshold=3

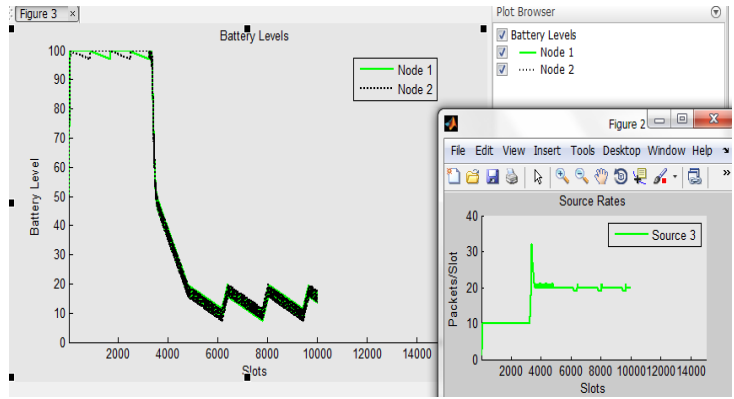

Threshold=5

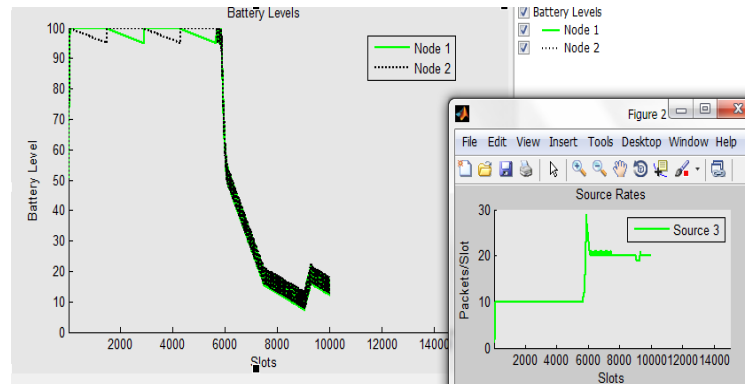

Graph of the total throughput and switches as a function of the threshold value:

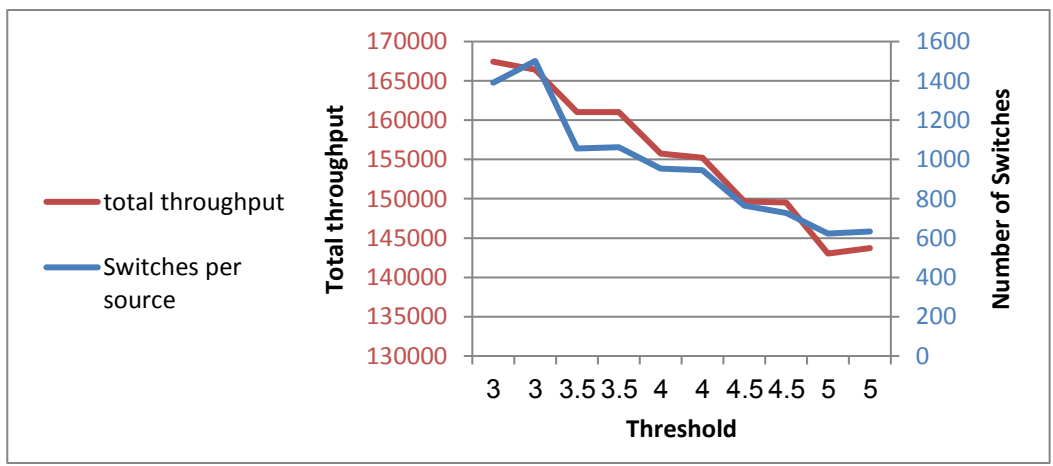

## Feedback with Danger Zones:

The value changes of the three tested parameters (Initial routing, Shift pause, Threshold) in this strategy generate the same effects as in the previous strategy.

**Allowed Drift-** It seem that if sharper slopes are allowed (higher Allowed drift values) the control reacts (by changing the transmission rate) slower. When the value was 0.3 (the highest value tested), the transmission rate did not converge to a stable optimal rate.

# Topology #2

Two sources, three intermediate nodes. The sources have one mutual path. There is a link that is not included in any path (NonPath link).

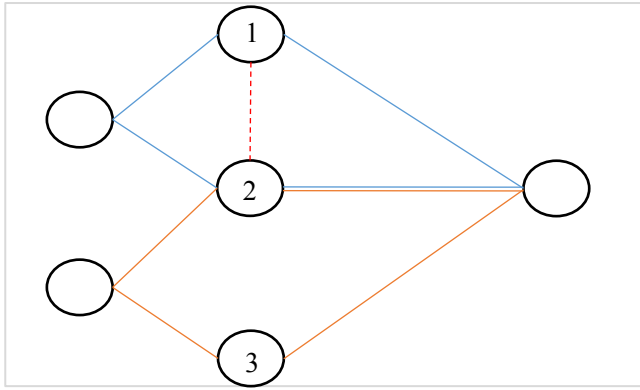

## Estimate:

**Initial Routing-** It seems that the topology is not stable. On different initial routings, the results were quite different, and even between both runs of the same initial routing there were sometimes significant differences. The different behaviors can be seen by looking at the difference in the Routing State Statistics, the switches count, and the battery level graphs. In most of the runs there was no fairness between the sources. The initial routing that had the best total throughput is [2,2] (where each source initially routes through its second path).

**Shift Pause-** as the value increases, the throughput decreases. When the initial routing is chosen as [2,2], the upper source makes a large amount of switches, and the lower source makes a small amount, but as the shift pause value gets higher, the number of switches in the upper source gets lower. Moreover, in that initial routing, more packets are sent from the upper source rather than from the lower source, and as the shift pause value gets higher, the number of packets sent from the upper lowers, and there is more fairness between the sources. Furthermore, for higher values, the control starts taking place later, and for the highest tested value, the control hasn't taken place at all.

**Thresholds-** as the value gets higher, fewer switches occur, and the throughput of the upper source decrease, while the throughput of the lower source remains the same for all values. Moreover, for higher thresholds values, the battery level graphs of nodes 1, 3 were less "wavy".

## Feedback with Danger Zones:

**Initial Routing-** as in the previous strategy, the topology isn't stable. The Routing State Statistics, the energy level graphs and the number of switches differ between runs. The initial routing [2, 2] had the best total throughput in this strategy as well.

**Shift Pause-** as the value gets higher, the throughput gets lower, fewer switches occur and the fairness gets better (the throughput of the upper source gets lower and more similar to the throughput of the lower source). As in the previous strategy, for higher values, it takes more time for the control to take place, even causing packet drops, and for the highest tested value, the control hasn't taken place at all.

**Thresholds-** as the value increases, less switches occur, and the throughput decreases. The upper source transmits more packets than the lower source, and changing the thresholds value has no significant effect on the fairness between the sources.

**Allowed drift-** lower values keep the battery level of the intermediate nodes from dropping to $B_{min}$. In high values, there were packet drops. As the value gets lower, the convergence of the transmission rates into optimal transmission rates take less time.

# Topology #3

This topology is constructed out of two sources and four intermediate nodes. The sources have one mutual path that is two nodes long. Node number 4 is in both paths of the lower source.

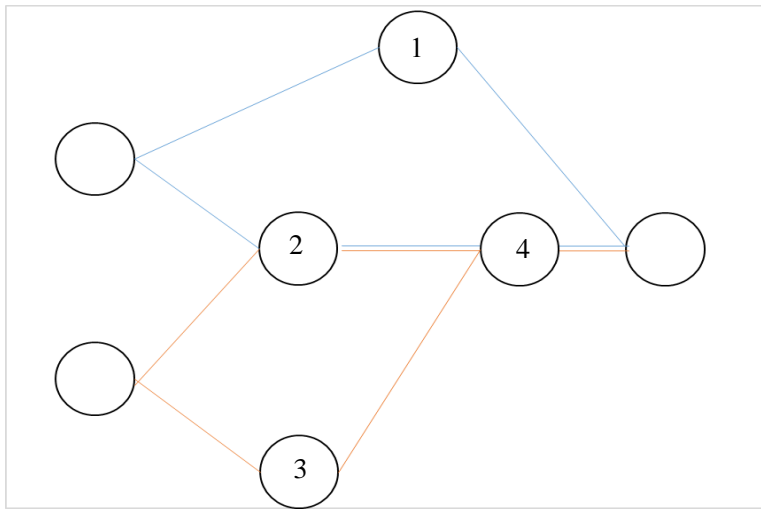

## Estimate:

**Initial Routing-** Changing the initial routing does not make significant effect on the output statistics or the output graphs. From the Routing State Statistics it is clear that the most common routing from the upper source is which it transmits through its upper path.

**Shift Pause-** as the value gets higher, the total throughput decreases, but the fairness gets better. However, comparing the results from the previous section (initial routings runs) where the shift pause parameters values are the default values of the simulation ([5,2,3]), with the results from this section (where all three shift pause parameters have the same value) on the run where the value is 3 ([3,3,3])- both had similar throughput and total switches number, but the fairness on the run with the default values (where the three parameters does not have equal values) has much better fairness and much less switches in the upper source.

**Thresholds-** as the value lowers, more switches occur, but there is no significant change in the total throughput.

## Feedback with Danger Zones:

**Initial Routing-** As in the previous strategy, there is no significant difference in results of different initial routings.

**Shift Pause-** as the value gets higher, the fairness between the sources gets better, but the total throughput gets slightly lower. Moreover, the total switch number is approximately the same in all shift pause values, but as the value gets higher, less of the switches occur at the higher source, and more of them occur at the lower source.

**Thresholds-** as the value lowers, the number of switches increase, especially the switches in the 2nd source, but there is no significant change in the total throughput.

**Allowed Drift-** as the parameter value increases, the slopes of the energy level graphs get sharper, and the energy levels change from high values to low values and vice versa rapidly.

In the lowest tested value, the energy level graphs were mostly balanced, while in the highest tested value, the energy level in some nodes dropped below $B_{min}$, and some packets were lost.

# Topology #4

Three sources, five intermediate nodes. Every source has an overlapping path with the source below and/or above it.

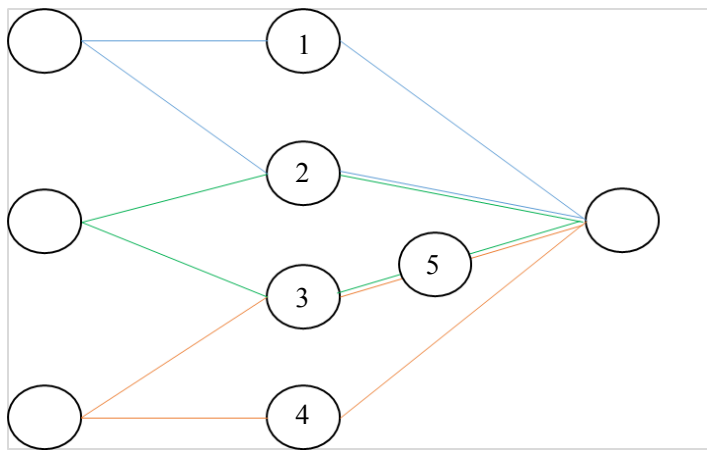

## Estimate:

**Initial Routing-** Changing the initial routing does not affect the throughput or any other statistics    From the Routing State Statistics we can see that most of the routings throughout the simulation are routings for which the 1st source transmits through its upper path and the 3rd source transmits through its lower path.

**Shift Pause-** as the value gets higher, less switches occur in the 2nd source (which is where almost all of the switches occur) and the total throughput decreases.  However, changing the shift pause value does not affect the Routing State Statistics.  As the tested shift pause value gets higher, the steeper the drop in the battery level graphs of nodes 2 and 3, and on the highest values, their energy levels drop to $B_{min}$.

**Thresholds-** as the value gets lower, more switches occur in the 2nd source, the total throughput increase, and the battery level graphs of nodes 2 and 3 get more "wavy".

Threshold=3

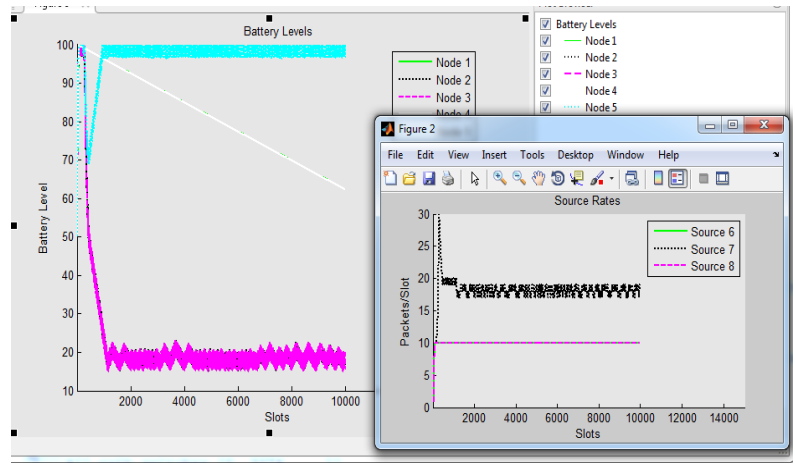

Threshold=4.5

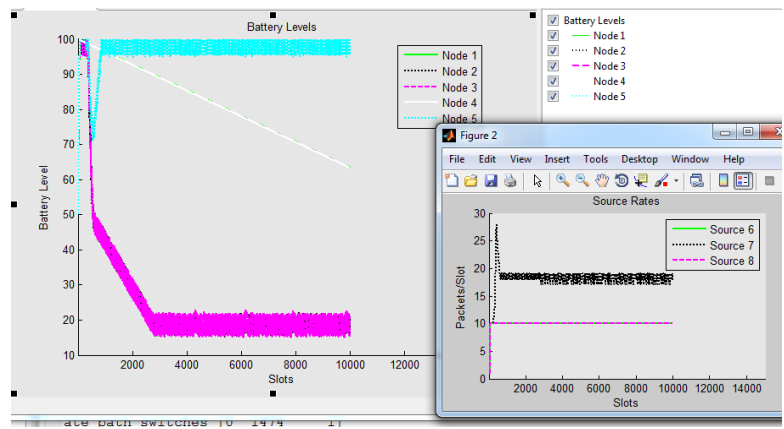

## Feedback with Danger Zones:

**Initial Routing-** As in the previous strategy, all of the initial routings resulted in similar Routing State Statistics, but in this strategy, the number of switches in the 1st and 3rd sources is higher than that in the previous strategy, and depends on the initial routing chosen.   Changing the values of the Shift Pause and Threshold parameters of this topology had the same effect as in the Estimates strategy.

**Allowed Drift-** as the value gets higher, it takes more time for the transmission rate of the 2nd source to get stable.  For the highest tested value, there were packet drops.

# Conclusions

## How the parameters affect the results

- As the threshold value gets higher, fewer switches occur throughout the run.
- Increasing the value of the shift pause parameters usually decreases the throughput, because it takes more time until the control takes place or reaches an optimal transmission rate.
- High Allowed_Drift values lead to sharper slopes of the energy level graphs that sometimes cause packets lost.  In addition, it seems that this value has an effect on the convergence of the transmission rates to an optimal rate. As for higher values, this convergence takes more time.
- Comparing the two strategies on the same topologies and parameters, over all the runs, the "Estimate" strategy almost always results in a better throughput.

# Strategies Comparison

In order to compare the three suggested strategies, we collected results of runs of all three strategies, on 4 different topologies. Runs on the same topology, have the same parameters, with the exception of the initial transmission rate, which in the feedback strategies is set to 1, and in the no feedback strategy is set to default value of 15 (or 19 on the 1st topology, using the analysis to get a rate that won't cause drops). The runs are 10000 slots long. The system and strategy parameters have the simulation system default values.

| scenario | strategy | Throughput per slot | Lost packets per slot |
|---|---|---|---|
| 1 source, 2 forwarding | NoFeedback | 19 | 0 |
|  | Feedback with Danger Zones | 15.5474 | 0 |
|  | Estimate | 15.727 | 0 |

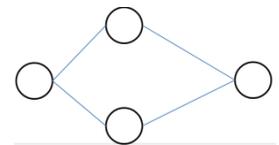

| scenario | strategy | Throughput per slot | Lost packets per slot |
|---|---|---|---|
| 2 sources, 3 forwarding | NoFeedback | 29.2045 | 0.7955 |
|  | Feedback with Danger Zones | 27.1856 | 0 |
|  | Estimate | 27.3143 | 0 |

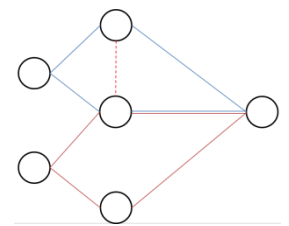

| scenario | strategy | Throughput per slot | Lost packets per slot |
|---|---|---|---|
| 2 sources, 4 forwarding | NoFeedback | 20.0216 | 9.9784 |
| | Feedback with Danger Zones | 19.8137 | 0 |
| | Estimate | 19.9066 | 0 |

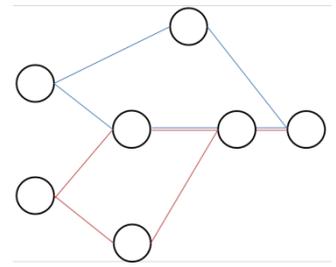

| scenario | strategy | Throughput per slot | Lost packets per slot |
|---|---|---|---|
| 3 sources, 5 forwarding | NoFeedback | 37.849 | 7.151 |
| | Feedback with Danger Zones | 37.7114 | 0 |
| | Estimate | 37.9375 | 0 |

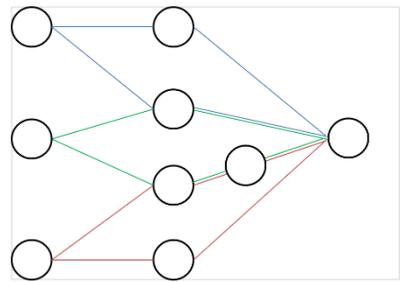

"No feedback" is good in the simple topology, where the transmission rate needed is simple to calculate, but in more complex topologies, finding the correct rate to avoid drops is much harder.  The feedback strategies avoid packet drops very well, while still maintaining a good throughput, even on more complex topologies.  The throughput per slot of the feedback strategies is usually lower than the throughput of the "No feedback" strategy, because the feedback strategies take some time to get a steady and high transmission rate, where on the "No feedback" strategy, the transmission rate is high from the first slot.    For all topologies, the results of the "Estimate" strategy are better than the results of the "Feedback with danger zones" strategy. This might indicate that a drift threshold that is estimated during the run is better than a constant threshold.

# Summary

In this project we have built a simulation with a convenient interface, so it can deal with the EnHANTs sensor network topologies we focused on. We have expanded the strategy that has been presented in the article into more complex strategies that include feedback properties. In these strategies the destination node can balance the transmission rate creating a better throughput with consideration to the topology and its parameters. These new strategies are more efficient than what was previously proposed.

# Future Improvements

In this project we focus on a specific case of the EnHANTs sensor networks.

In the future, the simulation can be expanded so it will include more cases:

a. Topologies with more than two routing paths per source.
b. Topologies with more than one destination node.
c. Sensor networks with source nodes that have a limited energy supply and an energy harvest ability (as the intermediate nodes)

For these extensions, it is required to expand the strategies, so they will fit the extended model. For example, the HDR strategy that we use is defined for two routing paths per source, but it is not obvious how to expand this logic for topologies that have more than two paths per source- should we have a threshold for every pair of paths or maybe use thresholds to indicate switching to the next path on a certain order of the paths. In addition, we might consider not pre-defining the routing paths and make the destination node to be able to determine the routing paths during the operation of the network. For that kind of networks, we need to develop logic for the destination node, which will make it able to decide on good routing paths, depending on the system current state.

# Appendix A: Simulation Results of the Parameter Examination Runs

## Topology #1

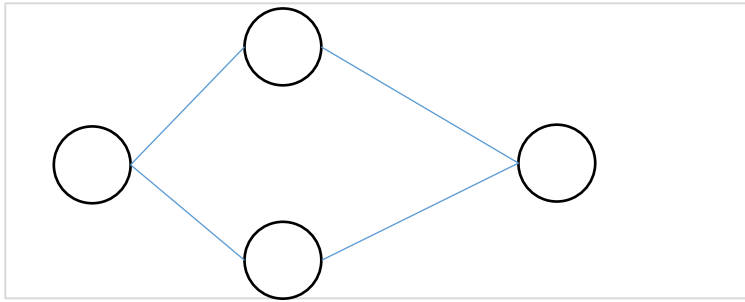

*Initial route:*

*Estimate:*

| Initial Route | Routing State Statistic | Switches per source | Packets sent by source | Packets at sink | Dropped Packets |
|---|---|---|---|---|---|
| 1 | 0.44 ; 0.56 | 757 | 157270 | 157270 | 0 |
| 1 | 0.43; 0.57 | 758 | 157365 | 157365 | 0 |
| 2 | 0.57; 0.43 | 753 | 157036 | 157036 | 0 |
| 2 | 0.56 ; 0.44 | 752 | 157226 | 157226 | 0 |

The chosen initial route is 1 (the upper path)

*Feedback with Danger Zones:*

| Initial Route | Routing State Statistic | Switches per source [s1] | Packets sent by source [s1] | Packets at sink [s1] | Dropped Packets [s1] |
|---|---|---|---|---|---|
| 1 | 0.44; 0.56 | 769 | 155474 | 155474 | 0 |
| 1 | 0.43; 0.57 | 765 | 154796 | 154796 | 0 |
| 2 | 0.57; 0.43 | 773 | 155438 | 155438 | 0 |
| 2 | 0.57; 0.43 | 770 | 155329 | 155329 | 0 |

## Shift Pause:

*Estimate:*

| Shift pause | Routing State Statistic | Switches per source | Packets sent by source | Packets at sink | Dropped Packets |
|---|---|---|---|---|---|
| 1 | 0.5; 0.5 | 1026 | 170947 | 170947 | 0 |
| 1 | 0.5; 0.6 | 963 | 171470 | 171470 | 0 |
| 2 | 0.43; 0.57 | 808 | 157219 | 157219 | 0 |
| 2 | 0.43; 0.57 | 770 | 156893 | 156893 | 0 |
| 3 | 0.5; 0.5 | 628 | 143383 | 143383 | 0 |
| 3 | 0.5; 0.5 | 631 | 143588 | 143588 | 0 |
| 5 | 0.5; 0.5 | 235 | 116126 | 116126 | 0 |
| 5 | 0.49; 0.51 | 235 | 116132 | 116132 | 0 |
| 8 | 0.44; 0.56 | 8 | 99149 | 99149 | 0 |
| 8 | 0.44; 0.56 | 8 | 99148 | 99148 | 0 |
| 15 | 0.44; 0.56 | 8 | 98889 | 98889 | 0 |
| 15 | 0.43; 0.57 | 8 | 98880 | 98880 | 0 |

*Feedback with Danger Zones:*

| Shift pause | Routing State Statistic | Switches per source | Packets sent by source | Packets at sink | Dropped Packets |
|---|---|---|---|---|---|
| 1 | 0.5; 0.5 | 939 | 168886 | 168886 | 0 |
| 1 | 0.5; 0.5 | 943 | 169211 | 169211 | 0 |
| 2 | 0.43; 0.57 | 814 | 155309 | 155309 | 0 |
| 2 | 0.43; 0.57 | 810 | 154997 | 154997 | 0 |
| 3 | 0.5; 0.5 | 589 | 141554 | 141554 | 0 |
| 3 | 0.5; 0.5 | 593 | 141764 | 141764 | 0 |
| 5 | 0.5; 0.5] | 220 | 115336 | 115336 | 0 |
| 5 | 0.5; 0.5 | 216 | 114961 | 114961 | 0 |
| 8 | 0.44; 0.56 | 8 | 99154 | 99154 | 0 |
| 8 | 0.44; 0.56 | 8 | 99134 | 99134 | 0 |
| 15 | 0.43; 0.57 | 8 | 98902 | 98902 | 0 |
| 15 | 0.43; 0.57 | 8 | 98908 | 98908 | 0 |

## Shift Pause:

## Threshold:

### Estimate:

| Threshold | Routing State Statistic | Switches per source | Packets sent by source | Packets at sink | Dropped Packets |
|---|---|---|---|---|---|
| 3 | 0.49; 0.51 | 1391 | 167436 | 167436 | 0 |
| 3 | 0.5; 0.5 | 1501 | 166427 | 166427 | 0 |
| 3.5 | 0.5; 0.5 | 1056 | 161015 | 161015 | 0 |
| 3.5 | 0.5; 0.5 | 1062 | 161023 | 161023 | 0 |
| 4 | 0.5; 0.5 | 953 | 155742 | 155742 | 0 |
| 4 | 0.5; 0.5 | 946 | 155235 | 155235 | 0 |
| 4.5 | 0.5; 0.5 | 765 | 149643 | 149643 | 0 |
| 4.5 | 0.5; 0.5 | 728 | 149548 | 149548 | 0 |
| 5 | 0.5; 0.5 | 623 | 143032 | 143032 | 0 |
| 5 | 0.5; 0.5 | 633 | 143718 | 143718 | 0 |

### Feedback with Danger Zones:

| Threshold | Routing State Statistic | Switches per source | Packets sent by source | Packets at sink | Dropped Packets |
|---|---|---|---|---|---|
| 3 | 0.5; 0.5 | 1462 | 164682 | 164682 | 0 |
| 3 | 0.5; 0.5 | 1480 | 165276 | 165276 | 0 |
| 3.5 | 0.5; 0.5 | 1125 | 159287 | 159287 | 0 |
| 3.5 | 0.5; 0.5 | 1134 | 158838 | 158838 | 0 |
| 4 | 0.5; 0.5 | 890 | 153353 | 153353 | 0 |
| 4 | 0.5; 0.5 | 893 | 153497 | 153497 | 0 |
| 4.5 | 0.5; 0.5 | 749 | 147535 | 147535 | 0 |
| 4.5 | 0.5; 0.5 | 752 | 147053 | 147053 | 0 |
| 5 | 0.5; 0.5 | 593 | 141694 | 141694 | 0 |
| 5 | 0.5; 0.5 | 597 | 141937 | 141937 | 0 |

## Allowed Drift:

### Feedback with Danger Zones:

| Allowed Drift | Routing State Statistic | Switches per source | Packets sent by source | Packets at sink | Dropped Packets |
|---|---|---|---|---|---|
| 0.3 | 0.5; 0.5 | 619 | 144322 | 143910 | 412 |
| 0.3 | 0.5; 0.5 | 629 | 144646 | 144235 | 411 |
| 0.03 | 0.5; 0.5 | 574 | 143300 | 143300 | 0 |
| 0.03 | 0.5; 0.5 | 573 | 143265 | 143265 | 0 |
| 0.003 | 0.5; 0.5 | 588 | 141324 | 141324 | 0 |
| 0.003 | 0.5; 0.5 | 596 | 141956 | 141956 | 0 |

# Topology #2

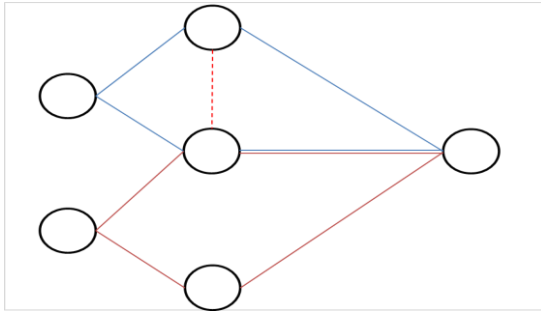

*Initial route:*

*Estimate:*

| Initial Route | Routing State Statistic [1 1; 1 2; 2 1; 2 2] | Switches per source [s1, s2] | Packets sent by source [s1, s2] | Packets at sink [s1, s2] | Dropped Packets | total throughput |
|---|---|---|---|---|---|---|
| 1,1 | [0.3; 0.63; 0.025; 0.052] | [190 912] | 102864 156742 | 102864 156742 | 0 | 259606 |
| 1,1 | [0.34; 0.64; 0.0035; 0.014] | [30 981] | 95141 163967 | 95141 163967 | 0 | 259108 |
| 1,2 | 0.0071; 0.5; 0.14; 0.34 | 948  34 | 159304 94085 | 159304 94085 | 0 | 253389 |
| 1,2 | 0.0054; 0.51; 0.14; 0.35 | 950  26 | 159811 93796 | 159811 93796 | 0 | 253607 |
| 2,1 | 0.0033; 0.61; 0.0032; 0.39 | 1040 3 | 173597 99375 | 173597 99375 | 0 | 272972 |
| 2,1 | 0.043; 0.71; 0.2; 0.045 | 871 877 | 134285 134823 | 134285 134823 | 0 | 269108 |
| 2,2 | 0.0034; 0.6; 0; 0.39 | 10432 | 173797 99346 | 173797 99346 | 0 | 273143 |
| 2,2 | 0.0034; 0.6; 0; 0.39 | 10432 | 173797 99346 | 173797 99346 | 0 | 273143 |

<mark>The chosen initial route is 2, 2 (the lower path for each source)</mark>

*Feedback with Danger Zones:*

| Initial Route | Routing State Statistic [1 1; 1 2; 2 1; 2 2] | Switches per source [s1, s2] | Packets sent by source [s1, s2] | Packets at sink [s1, s2] | Dropped Packets [s1, s2] | total throughput |
|---|---|---|---|---|---|---|
| 1,1 | 0.28; 0.63; 0.038; 0.045 | 224 909 | 103181 155127 | 103181 155127 | 0 | 258308 |
| 1,1 | 0.13; 0.65; 0.13; 0.091 | 686 813 | 126838 136212 | 126838 136212 | 0 | 263050 |
| 1,2 | 0.13; 0.49; 0.22; 0.15 | 684 658 | 124756 123079 | 124756 123079 | 0 | 247835 |
| 1,2 | 0.18; 0.43; 0.15; 0.23 | 572 487 | 125005 119127 | 125005 119127 | 0 | 244132 |
| 2,1 | 0.18; 0.58; | 589 589 | 129670 | 129670 | 0 | 259698 |

| | 0.045; 0.19 | | 130028 | 130028 | | |
|---|---|---|---|---|---|---|
| 2,1 | 0.0034; 0.6; 0.0057; 0.39 | 1084   13 | 172521 99433 | 172521 99433 | 0 | 271954 |
| 2,2 | 0.045; 0.62; 0.017; 0.32 | 896   152 | 157978 105789 | 157978 105789 | 0 | 263767 |
| 2,2 | 0.0037; 0.6; 0.0034; 0.39 | 1085   16 | 172364 99492 | 172364 99492 | 0 | 271856 |

## Shift Pause:

*Estimate:*

| Shift pause values | Routing State Statistic | Switches per source [s1, s2] | Packets sent by source [s1, s2] | Packets at sink [s1, s2] | Dropped Packets | total throughput |
|---|---|---|---|---|---|---|
| 1 | 0.0033; 0.54; 0; 0.46 | 1253   2 | 187009 99424 | 187009 99424 | 0 | 286433 |
| 1 | 0.0033; 0.54; 0; 0.46 | 1250   2 | 86939 99424 | 86939 99424 | 0 | 286363 |
| 2 | 0.019; 0.62; 0.0089; 0.36 | 983   68 | 167658 102661 | 167658 102661 | 0 | 270319 |
| 2 | 0.014; 0.62; 0.0095; 0.36 | 988   60 | 168329 102108 | 168329 102108 | 0 | 270437 |
| 3 | 0.0034; 0.6; 0; 0.39 | 1036   2 | 172967 99350 | 172967 99350 | 0 | 272317 |
| 3 | 0.0067; 0.64; 0; 0.36 | 953   4 | 167047 99349 | 167047 99349 | 0 | 266396 |
| 5 | 0.0068; 0.63; 0; 0.36 | 927   4 | 165002 99275 | 165002 99275 | 0 | 264277 |
| 5 | 0.01; 0.7; 0; 0.29 | 754   6 | 152739 99263 | 152739 99263 | 0 | 252002 |
| 8 | 0.014; 0.75; 0; 0.23 | 581   8 | 140072 99159 | 139746 99159 | 362, 0 | 239231 |
| 8 | 0.013; 0.75; 0; 0.23 | 579   8 | 39938 99153 | 139622 99153 | 316,0 | 239091 |
| 15 | 0.02; 0.95; 0; 0.031 | 13   12 | 98886 98915 | 98886 98915 | 0 | 197801 |
| 15 | 0.02; 0.94; 0; 0.042 | 15   12 | 98876 98911 | 98876 98911 | 0 | 197787 |

*Feedback with Danger Zones:*

| Shift pause values | Routing State Statistic [1 1; 1 2; 2 1; 2 2] | Switches per source [s1, s2] | Packets sent by source [s1, s2] | Packets at sink [s1, s2] | Dropped Packets [s1, s2] | total throughput |
|---|---|---|---|---|---|---|
| 1 | 0.0048; 0.54; 0.0062; 0.45 | 1237   30 | 184735 99940 | 184735 99940 | 0 | 284675 |
| 1 | 0.0041; 0.54; 0.0048; 0.45 | 1235   22 | 184924 99732 | 184924 99732 | 0 | 284656 |
| 2 | 0.021; | 985   116 | 164497 | 164497 | 0 | 267609 |

|   | 0.62; 0.02; 0.34 |   | 103112 | 103112 |   |   |
|---|---|---|---|---|---|---|
| 2 | 0.021; 0.62; 0.018; 0.34 | 986  108 | 164764 102895 | 164764 102895 | 0 | 267659 |
| 3 | 0.0068; 0.64; 0; 0.36 | 976  4 | 165296 99351 | 165296 99351 | 0 | 264647 |
| 3 | 0.0037; 0.6; 0.0024; 0.39 | 1066  12 | 171156 99465 | 171156 99465 | 0 | 270621 |
| 5 | 0.01; 0.7; 0; 0.29 | 777  6 | 152054 99274 | 152054 99274 | 0 | 251328 |
| 5 | 0.0067; 0.63; 0; 0.36 | 961  4 | 164348 99278 | 164348 99278 | 0 | 263626 |
| 8 | 0.014; 0.76; 0; 0.23 | 566  8 | 137754 99171 | 137436 99171 | 318 | 236925, 236607 |
| 8 | 0.077; 0.79; 0.0079; 0.12 | 277  178 | 115376 108092 | 115270 108092 | 106 | 223468, 223362 |
| 15 | 0.02; 0.94; 0; 0.042 | 15  12 | 98894 98911 | 98894 98911 | 0 | 197805 |
| 15 | 0.02; 0.95; 0; 0.031 | 13  12 | 98887 98916 | 98887 98916 | 0 | 197803 |

*Threshold:*

*Estimate:*

| Threshold | Routing State Statistic | Switches per source [s1, s2] | Packets sent by source [s1, s2] | Packets at sink [s1, s2] | Dropped Packets | total throughput |
|---|---|---|---|---|---|---|
| 3 | 0.0084; 0.54; 0; 0.45 | 1936  4 | 185935 99300 | 185935 99300 | 0 | 285235 |
| 3 | 0.0084; 0.54; 0; 0.45 | 1935  4 | 185726 99304 | 185726 99304 | 0 | 285030 |
| 3.5 | 0.0024; 0.57; 0; 0.42 | 1549  2 | 179894 99350 | 179894 99350 | 0 | 279244 |
| 3.5 | 0.0024; 0.57; 0; 0.42 | 1546  2 | 180028 99350 | 180028 99350 | 0 | 279378 |
| 4 | 0.0027; 0.61; 0.0005; 0.38 | [1273  4 | 173046 99347 | 173046 99347 | 0 | 272393 |
| 4 | 0.0027; 0.59; 0; 0.41 | 1344  2 | 177377 99351 | 177377 99351 | 0 | 276728 |
| 4.5 | 0.0061; 0.62; 0; 0.37 | 1110  4 | 169855 99343 | 169855 99343 | 0 | 269198 |
| 4.5 | 0.0061; 0.63; 0; 0.37 | 1107  4 | 169716 99351 | 169716 99351 | 0 | 269067 |
| 5 | 0.0068; 0.64; 0; 0.36 | 948  4 | 166659 99346 | 166659 99346 | 0 | 266005 |
| 5 | [0.0034; 0.6; 0; 0.4 | 1042  2 | 173321 99342 | 173321 99342 | 0 | 272663 |

*Feedback with Danger Zones:*

| Threshold | Routing State Statistic [1 1; 1 2; 2 1; 2 2] | Switches per source [s1, s2] | Packets sent by source [s1, s2] | Packets at sink [s1, s2] | Dropped Packets [s1, s2] | total throughput |
|---|---|---|---|---|---|---|
| 3 | 0.0085; 0.54; 0; 0.45 | 1897   4 | 184802 99293 | 184802 99293 | 0 | 284095 |
| 3 | 0.0081; 0.54; 0; 0.45 | 1895   4 | 184774 99306 | 184774 99306 | 0 | 284080 |
| 3.5 | 0.0024; 0.57; 0; 0.43 | 1581   2 | 180010 99348 | 180010 99348 | 0 | 279358 |
| 3.5 | 0.0047; 0.6; 0; 0.4 | 1480   4 | 174874 99350 | 174874 99350 | 0 | 274224 |
| 4 | 0.0053; 0.61; 0; 0.38 | 1293   4 | 172333 99349 | 172333 99349 | 0 | 271682 |
| 4 | 0.0054; 0.61; 0; 0.38 | 1292   4 | 172282 99347 | 172282 99347 | 0 | 271629 |
| 4.5 | 0.0062; 0.63; 0; 0.37 | 1133   4 | 168524 99348 | 168524 99348 | 0 | 267872 |
| 4.5 | 0.0061; 0.63; 0; 0.37 | 1132   4 | 168502 99347 | 168502 99347 | 0 | 267849 |
| 5 | 0.0068; 0.64; 0.0005; 0.35 | 972   6 | 165028 99364 | 165028 99364 | 0 | 264392 |
| 5 | 0.0033; 0.6; 0.002; 0.39 | 1069   10 | 171420 99464 | 171420 99464 | 0 | 270884 |

*Allowed Drift:*

*Feedback with Danger Zones:*

| Allowed Drift | Routing State Statistic [1 1; 1 2; 2 1; 2 2] | Switches per source [s1, s2] | Packets sent by source [s1, s2] | Packets at sink [s1, s2] | Dropped Packets [s1, s2] | total throughput |
|---|---|---|---|---|---|---|
| 0.3 | 0.0032; 0.61; 0; 0.39 | 1066   2 | 173205 99349 | 172668 99349 | 537   0 | 272017 |
| 0.3 | 0.0068; 0.64; 0; 0.35 | 975   4 | 167273 99350 | 166735 99350 | 538   0 | 266085 |
| 0.03 | 0.0066; 0.64; 0; 0.35 | 946   4 | 166519 99349 | 166519 99349 | 0 | 265868 |
| 0.03 | 0.0068; 0.64; 0; 0.35 | 945   4 | 166445 99348 | 166445 99348 | 0 | 265793 |
| 0.003 | 0.0037; 0.6; 0.0024; 0.39 | 1072   12 | 171540 99463 | 171540 99463 | 0 | 271003 |
| 0.003 | 0.0066; 0.64; 0; 0.36 | 982   5 | 165696 99344 | 165696 99344 | 0 | 265040 |

# Topology #3

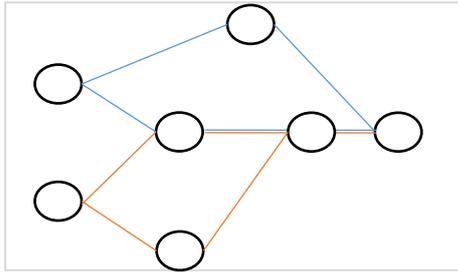

*Initial route:*

*Estimate:*

| Initial Route | Routing State Statistic [1 1; 1 2; 2 1; 2 2] | Switches per source [s1, s2] | Packets sent by source [s1, s2] | Packets at sink [s1, s2] | Dropped Packets | total throughput |
|---|---|---|---|---|---|---|
| 1,1 | 0.34; 0.65; 0; 0.0007 | 2  209 | 99344  99722 | 99344 99722 | 0 | 199066 |
| 1,1 | 0.34; 0.66; 0; 0.0007 | 2  209 | 99347  99722 | 99347 99722 | 0 | 199069 |
| 1,2 | 0.36; 0.64; 0; 0.0007 | 2  211 | 99349  99652 | 99349 99652 | 0 | 199001 |
| 1,2 | 0.35; 0.65; 0; 0.0007 | 2  209 | 99347  99604 | 99347 99604 | 0 | 198951 |
| 2,1 | 0.34; 0.66; 0.0029; 0.0007 | 3  209 | 99377  99685 | 99377 99685 | 0 | 199062 |
| 2,1 | 0.34; 0.66; 0.0029; 0.0007 | 3  210 | 99378  99690 | 99378 99690 | 0 | 199068 |
| 2,2 | 0.34; 0.66; 0; 0.0039 | 3  209 | 99375  99683 | 99375 99683 | 0 | 199058 |
| 2,2 | 0.34; 0.65; 0; 0.0039 | 3  209 | 99379  99677 | 99379 99677 | 0 | 199056 |

==The chosen initial route is 1,1 (the upper path for each source)==

*Feedback with Danger Zones:*

| Initial Route | Routing State Statistic | | | | Switches per source | Packets sent by source | Packets at sink | Dropped Packets | total throughput |
|---|---|---|---|---|---|---|---|---|---|
| | 2 2 | 2 1 | 1 2 | 1 1 | [s1,s2] | [s1, s2] | [s1, s2] | | |
| 1,1 | 0.009 | 0.000 | 0.690 | 0.300 | 26  177 | 100002 98046 | 100002 98046 | 0 | 198048 |
| 1,1 | 0.012 | 0.000 | 0.700 | 0.290 | 34  171 | 100351 97757 | 100351 97757 | 0 | 198108 |
| 1,2 | 0.013 | 0.000 | 0.710 | 0.280 | 36  166 | 100434 97553 | 100434 97553 | 0 | 197987 |
| 1,2 | 0.010 | 0.000 | 0.700 | 0.290 | 30  174 | 100221 97846 | 100221 97846 | 0 | 198067 |
| 2,1 | 0.010 | 0.003 | 0.690 | 0.290 | 29  177 | 100161 98005 | 100161 98005 | 0 | 198166 |
| 2,1 | 0.010 | 0.003 | 0.690 | 0.300 | 31  177 | 100220 97951 | 100220 97951 | 0 | 198171 |

| 2,2 | 0.013 | 0.000 | 0.700 | 0.290 | 29 176 | 100115 98022 | 100115 98022 | 0 | 198137 |
| 2,2 | 0.014 | 0.001 | 0.710 | 0.280 | 33 171 | 100286 97797 | 100286 97797 | 0 | 198083 |

## Shift Pause:

*Estimate:*

| Shift pause | Routing State Statistic [1 1; 1 2; 2 1; 2 2] | | | | Switches per source [s1, s2] | total switches | Packets sent by source [s1, s2] | | Packets at sink [S1,S2] | | Dropped Packets | Total throughput |
|---|---|---|---|---|---|---|---|---|---|---|---|---|
| 1 | 0.19 | 0.66 | 0.0004 | 0.14 | 421 89 | 510 | 124674 | 74550 | 74550 | 124674 | 0 | 199224 |
| 1 | 0.19 | 0.66 | 0.0004 | 0.14 | 422 89 | 511 | 124755 | 74460 | 74460 | 124755 | 0 | 199215 |
| 2 | 0.3 | 0.69 | 0 | 0.012 | 36 177 | 213 | 100738 | 98007 | 98007 | 100738 | 0 | 198745 |
| 2 | 0.3 | 0.69 | 0 | 0.012 | 36 177 | 213 | 100716 | 98029 | 98029 | 100716 | 0 | 198745 |
| 3 | 0.24 | 0.74 | 0 | 0.022 | 68 153 | 221 | 102094 | 96952 | 96952 | 102094 | 0 | 199046 |
| 3 | 0.24 | 0.73 | 0 | 0.023 | 70 153 | 223 | 102166 | 96912 | 96912 | 102166 | 0 | 199078 |
| 5 | 0.3 | 0.7 | 0 | 0.0042 | 12 187 | 199 | 99432 | 98585 | 98585 | 99432 | 0 | 198017 |
| 5 | 0.3 | 0.7 | 0 | 0.0042 | 12 187 | 199 | 99433 | 98593 | 98593 | 99433 | 0 | 198026 |
| 8 | 0.29 | 0.71 | 0 | 0.0021 | 6 189 | 195 | 99159 | 98721 | 98721 | 99159 | 0 | 197880 |
| 8 | 0.28 | 0.71 | 0 | 0.0021 | 6 187 | 193 | 99138 | 98676 | 98676 | 99138 | 0 | 197814 |
| 15 | 0.32 | 0.68 | 0 | 0 | 0 203 | 203 | 98914 | 99221 | 99221 | 98914 | 0 | 198135 |
| 15 | 0.32 | 0.68 | 0 | 0 | 0 204 | 204 | 98907 | 99230 | 99230 | 98907 | 0 | 198137 |

*Feedback with Danger Zones:*

| Shift pause values | Routing State Statistic [1 1; 1 2; 2 1; 2 2] | Switches per source [s1, s2] | Packets sent by source [s1, s2] | Packets at sink [s1, s2] | Dropped Packets | total throughput |
|---|---|---|---|---|---|---|
| 1 | 0.31; 0.67; 0.003; 0.015 | 43 169 | 100709 97454 | 100709 97454 | 0 | 198163 |
| 1 | 0.31; 0.67; 0.003; 0.015 | 43 169 | 100709 97455 | 100709 97455 | 0 | 198164 |
| 2 | 0.3; 0.69; 0.0029; 0.014 | 41 169 | 100628 97532 | 100628 97532 | 0 | 198160 |
| 2 | 0.29; 0.69; 0.0029; 0.017 | 49 162 | 100929 97226 | 100929 97226 | 0 | 198155 |
| 3 | 0.28; 0.71; 0.0029; 0.007 | 21 179 | 99851 98234 | 99851 98234 | 0 | 198085 |
| 3 | 0.28; 0.71; 0.0029; 0.007 | 21 179 | 99849 98248 | 99849 98248 | 0 | 198097 |
| 5 | 0.36; 0.64; 0.0031; 0.0033 | 11 194 | 99370 98533 | 99370 98533 | 0 | 197903 |
| 5 | 0.41; 0.58; 0.0044; 0.0055 | 21 198 | 99708 98310 | 99708 98310 | 0 | 198018 |
| 8 | 0.29; 0.7; 0.0046; 0.0011 | 9 190 | 99195 98530 | 99195 98530 | 0 | 197725 |
| 8 | 0.29; 0.71; 0.0045; 0.0012 | 9 190 | 99186 98516 | 99186 98516 | 0 | 197702 |
| 15 | 0.3; 0.7; 0.0029; 0.0028 | 9 185 | 98925 98278 | 98925 98278 | 0 | 197203 |
| 15 | 0.33; 0.67; 0.0034; 0.0002 | 3 199 | 98939 98738 | 98939 98738 | 0 | 197677 |

## Threshold:

### Estimate:

| Threshold | Routing State Statistic [1 1; 1 2; 2 1; 2 2] | Switches per source [s1,s2,s3] | Packets sent by source [s1, s2, s3] | Packets at sink [s1, s2, s3] | Dropped Packets | total throughput |
|---|---|---|---|---|---|---|
| 3 | 0.2; 0.77; 0; 0.032 | 160 223 | 103535 96144 | 103535 96144 | 0 | 199679 |
| 3 | 0.28; 0.71; 0; 0.016 | 79 283 | 101336 97696 | 101336 97696 | 0 | 199032 |
| 3.5 | 0.27; 0.71; 0; 0.017 | 69 237 | 101481 97532 | 101481 97532 | 0 | 199013 |
| 3.5 | 0.27; 0.71; 0; 0.018 | 70 237 | 101527 97517 | 101527 97517 | 0 | 199044 |
| 4 | 0.24; 0.74; 0; 0.022 | 78 193 | 102005 97075 | 102005 97075 | 0 | 199080 |
| 4 | 0.24; 0.73; 0; 0.022 | 78 195 | 101979 97085 | 101979 97085 | 0 | 199064 |
| 4.5 | 0.24; 0.74; 0; 0.023 | 78 169 | 102248 96947 | 102248 96947 | 0 | 199195 |
| 4.5 | 0.24; 0.74; 0; 0.023 | 77 169 | 102244 96946 | 102244 96946 | 0 | 199190 |
| 5 | 0.24; 0.74; 0; 0.022 | 68 153 | 102092 96956 | 102092 96956 | 0 | 199048 |
| 5 | 0.24; 0.74; 0; 0.024 | 72 151 | 102237 96855 | 102237 96855 | 0 | 199092 |

### Feedback with Danger Zones:

| Threshold | Routing State Statistic [1 1; 1 2; 2 1; 2 2] | Switches per source [s1, s2] | Packets sent by source [s1, s2] | Packets at sink [s1, s2] | Dropped Packets | Total throughput |
|---|---|---|---|---|---|---|
| 3 | 0.33; 0.66; 0.0018; 0.0068 | 35 311 | 99793 98263 | 99793 98263 | 0 | 198056 |
| 3 | 0.32; 0.67; 0.0018; 0.0048 | 25 314 | 99590 98483 | 99590 98483 | 0 | 198073 |
| 3.5 | 0.34; 0.65; 0.0022; 0.0069 | 29 267 | 99816 98233 | 99816 98233 | 0 | 198049 |
| 3.5 | 0.34; 0.65; 0.0028; 0.0073 | 33 265 | 99926 98137 | 99926 98137 | 0 | 198063 |
| 4 | 0.29; 0.7; 0.0023; 0.0072 | 25 225 | 99839 98210 | 99839 98210 | 0 | 198049 |
| 4 | 0.29; 0.7; 0.0023; 0.0072 | 25 225 | 99836 98220 | 99836 98220 | 0 | 198056 |
| 4.5 | 0.27; 0.72; 0.0033; 0.01 | 37 192 | 100284 97837 | 100284 97837 | 0 | 198121 |
| 4.5 | 0.28; 0.71; 0.0026; 0.008 | 29 197 | 99981 98106 | 99981 98106 | 0 | 198087 |
| 5 | 0.29; 0.7; 0.0029; 0.007 | 21 179 | 99847 98246 | 99847 98246 | 0 | 198093 |
| 5 | 0.29; 0.7; 0.0029; 0.007 | 21 180 | 99854 98242 | 99854 98242 | 0 | 198096 |

## Allowed Drift:

*Feedback with Danger Zones:*

| Allowed Drift | Routing State Statistic [1 1; 1 2; 2 1; 2 2] | Switches per source [s1, s2] | Packets sent by source [s1, s2] | Packets at sink [s1, s2] | Dropped Packets [s1, s2] | total throughput |
|---|---|---|---|---|---|---|
| 0.3 | 0.13; 0.79; 0.0029; 0.079 | 227  141 | 114617 84917 | 114588 84917 | 29  0 | 199505 |
| 0.3 | 0.13; 0.79; 0.0029; 0.079 | 228  141 | 114681 84874 | 114652 84874 | 29  0 | 199526 |
| 0.03 | 0.26; 0.73; 0.0029; 0.0041 | 13  185 | 99697  98672 | 99697  98672 | 0 | 198369 |
| 0.03 | 0.25; 0.75; 0.0029; 0.0045 | 14  181 | 99711  98649 | 99711  98649 | 0 | 198360 |
| 0.003 | 0.29; 0.7; 0.0029; 0.007 | 21  179 | 99850  98234 | 99850  98234 | 0 | 198084 |
| 0.003 | 0.28; 0.71; 0.0029; 0.0083 | 25  177 | 99982  98099 | 99982  98099 | 0 | 198081 |

# Topology #4

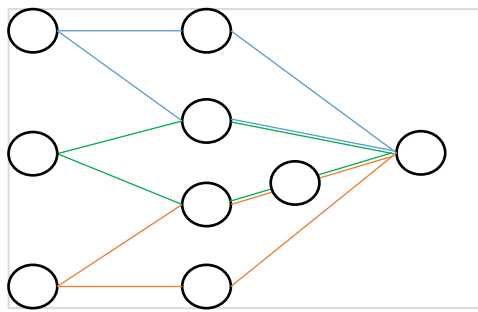

## Initial route:

*Estimate:*

| Initial Route | Routing State Statistic [1 1 1; 1 1 2; 1 2 1; 1 2 2; 2 1 1; 2 1 2; 2 2 1; 2 2 2] | Switches per source [s1, s2, s3] | Packets sent by source [s1, s2, s3] | Packets at sink [s1, s2, s3] | Dropped Packets [s1, s2, s3] | total throughput |
|---|---|---|---|---|---|---|
| 1 1 1 | 0.0053; 0.52; 0; 0.47; 0; 0; 0; 0 | 0  1347  1 | 99348 180226 99361 | 99348 180226 99361 | 0 | 378935 |
| 1 1 1 | 0.0053; 0.52; 0; 0.47; 0; 0; 0; 0 | 0  1347  1 | 99348 180226 99361 | 99348 180226 99361 | 0 | 378935 |
| 1 1 2 | 0; 0.52; 0; 0.48; 0; 0; 0; 0 | 0  1350  0 | 9345 180308 99350 | 9345 180308 99350 | 0 | 379003 |
| 1 1 2 | 0; 0.52; 0; 0.48; 0; 0; 0; 0 | 0  1349  0 | 99350 180254 99346 | 99350 180254 99346 | 0 | 378950 |
| 1 2 1 | 0; 0.53; 0.0029; | 0  1353  1 | 99351 | 99351 | 0 | 379375 |

| | Routing State Statistic [1 1 1; 1 1 2; 1 2 1; 1 2 2; 2 1 1; 2 1 2; 2 2 1; 2 2 2] | Switches per source [s1, s2, s3] | Packets sent by source [s1, s2, s3] | Packets at sink [s1, s2, s3] | Dropped Packets [s1, s2, s3] | total throughput |
|---|---|---|---|---|---|---|
| | 0.47; 0; 0; 0; 0 | | 180648 99376 | 180648 99376 | | |
| 1 2 1 | 0; 0.53; 0.0029; 0.47; 0; 0; 0; 0 | 0  1353   1 | 99346 180639 99376 | 99346 180639 99376 | 0 | 379361 |
| 1 2 2 | 0; 0.52; 0; 0.48; 0; 0; 0; 0 | 0  1355   0 | 99351 180613 99347 | 99351 180613 99347 | 0 | 379311 |
| 1 2 2 | 0; 0.52; 0; 0.48; 0; 0; 0; 0 | 0  1354   0 | 99348 180588 99352 | 99348 180588 99352 | 0 | 379288 |
| 2 1 1 | 0.0021; 0.52; 0; 0.47; 0.0032; 0; 0; 0 | 1  1346   1 | 99374 180199 99370 | 99374 180199 99370 | 0 | 378943 |
| 2 1 1 | 0.0021; 0.52; 0; 0.47; 0.0032; 0; 0; 0 | 1  1344   1 | 99377 180169 99363 | 99377 180169 99363 | 0 | 378909 |
| 2 1 2 | 0; 0.52; 0; 0.48; 0; 0.0032; 0; 0 | 1  1351   0 | 99369 180436 99345 | 99369 180436 99345 | 0 | 379150 |
| 2 1 2 | 0; 0.52; 0; 0.48; 0; 0.0032; 0; 0 | 1  1352   0 | 99379 180436 99347 | 99379 180436 99347 | 0 | 379162 |
| 2 2 1 | 0; 0.52; 0; 0.47; 0; 0; 0.0029; 0.0029 | 1  1352   1 | 99325 180467 99374 | 99325 180467 99374 | 0 | 379166 |
| 2 2 1 | 0; 0.52; 0; 0.47; 0; 0; 0.0029; 0.0029 | 1  1351   1 | 99328 180410 99374 | 99328 180410 99374 | 0 | 379112 |
| 2 2 2 | 0; 0.52; 0; 0.48; 0; 0; 0; 0.0058 | 1  1343   0 | 99334 179860 99348 | 99334 179860 99348 | 0 | 378542 |
| 2 2 2 | 0; 0.52; 0; 0.48; 0; 0; 0; 0.0058 | 1  1342   0 | 99313 179880 99343 | 99313 179880 99343 | 0 | 378536 |

==The chosen initial route is 1,2,1==

*Feedback with Danger Zones:*

| Initial Route | Routing State Statistic [1 1 1; 1 1 2; 1 2 1; 1 2 2; 2 1 1; 2 1 2; 2 2 1; 2 2 2] | Switches per source [s1, s2, s3] | Packets sent by source [s1, s2, s3] | Packets at sink [s1, s2, s3] | Dropped Packets [s1, s2, s3] | total throughput |
|---|---|---|---|---|---|---|
| 1 1 1 | 0.0058; 0.52; 0.0041; 0.47; 0; 0.004; 0; 0 | 16  1322  19 | 99466 177763 99515 | 99466 177763 99515 | 0 | 376744 |
| 1 1 1 | 0.0059; 0.52; 0.0034; 0.47; 0; 0.0039; 0; 0.0003 | 16  1321  17 | 99504 177799 99489 | 99504 177799 99489 | 0 | 376792 |
| 1 1 2 | 0.013; 0.51; 0.002; 0.47; 0.0002; 0.0036; 0; 0.0009 | 16  1297  38 | 99495 176017 101068 | 99495 176017 101068 | 0 | 376580 |
| 1 1 2 | 0.0012; 0.51; 0.0048; 0.48; 0; 0.0045; 0; 0 | 18  1323  22 | 99491 177646 99690 | 99491 177646 99690 | 0 | 376827 |
| 1 2 1 | 0.0054; 0.52; 0.0092; 0.46; 0; 0.0039; 0; | 16  1319  39 | 99491 177200 100423 | 99491 177200 100423 | 0 | 377114 |
| | 0.0002 | | | | | |
| 1 2 1 | 0.0001; 0.52; 0.0068; 0.47; 0; 0.0039; 0; 0.0003 | 16 1330 17 | 99493 178261 99495 | 99493 178261 99495 | 0 | 377249 |
| 1 2 2 | 0.0006; 0.52; 0.0038; 0.47; 0; 0.0039; 0; 0.0003 | 16 1326 16 | 99501 178177 99481 | 99501 178177 99481 | 0 | 377159 |
| 1 2 2 | 0.044; 0.5; 0.0074; 0.44; 0.0004; 0.0047; 0; 0.001 | 20 1227 134 | 99663 170571 106360 | 99663 170571 106360 | 0 | 376594 |
| 2 1 1 | 0.0025; 0.52; 0.0023; 0.47; 0.0032; 0.002; 0; 0.0001 | 9 1323 11 | 99400 178315 99435 | 99400 178315 99435 | 0 | 377150 |
| 2 1 1 | 0.0024; 0.52; 0.0035; 0.47; 0.0032; 0.004; 0; 0.0001 | 17 1323 17 | 99514 177818 99468 | 99514 177818 99468 | 0 | 376800 |
| 2 1 2 | 0.01; 0.46; 0.0044; 0.46; 0.0029; 0.018; 0.0011; 0.049 | 179 1203 48 | 108534 166356 101220 | 108534 166356 101220 | 0 | 376110 |
| 2 1 2 | 0.042; 0.49; 0.005; 0.45; 0.0003; 0.0077; 0; 0.007 | 31 1226 120 | 100312 170425 105508 | 100312 170425 105508 | 0 | 376245 |
| 2 2 1 | 0.0002; 0.52; 0.0038; 0.47; 0; 0.0029; 0.0029; 0.0032 | 13 1326 17 | 99385 178121 99496 | 99385 178121 99496 | 0 | 377002 |
| 2 2 1 | 0.0007; 0.51; 0.0027; 0.47; 0; 0.004; 0.0029; 0.0031 | 17 1327 13 | 99445 178174 99452 | 99445 178174 99452 | 0 | 377071 |
| 2 2 2 | 0.0008; 0.51; 0.0035; 0.48; 0; 0.0041; 0; 0.0058 | 17 1316 16 | 99460 177500 99466 | 99460 177500 99466 | 0 | 376426 |
| 2 2 2 | 0.0003; 0.51; 0.0037; 0.48; 0; 0.003; 0; 0.006 | 13 1319 16 | 99386 177583 99474 | 99386 177583 99474 | 0 | 376443 |

## Shift Pause:

*Estimate:*

| Shift pause | Routing State Statistic | Switches per source [s1, s2, s3] | Packets sent by source [s1, s2, s3] | Packets at sink [s1, s2, s3] | Dropped Packets | total throughput |
|---|---|---|---|---|---|---|
| 1 | 0; 0.52; 0.0029; 0.48; 0; 0; 0 | 0 1385 1 | 99424 181784 99451 | 99424 181784 99451 | 0 | 380659 |
| 1 | 0; 0.52; 0.0029; 0.48; 0; 0; 0; 0 | 0 1384 1 | 99424 181796 99451 | 99424 181796 99451 | 0 | 380671 |
| 2 | 0; 0.52; 0.0029; 0.47; 0; 0; 0; 0 | 0 1363 1 | 99388 180932 99415 | 99388 180932 99415 | 0 | 379735 |
| 2 | 0; 0.52; 0.0029; 0.47; 0; 0; 0; 0 | 0 1369 1 | 99388 180942 99415 | 99388 180942 99415 | 0 | 379745 |
| 3 | 0; 0.53; 0.0029; 0.47; 0; 0; 0; 0 | 0 1360 1 | 99352 180261 99376 | 99352 180261 99376 | 0 | 378989 |
| 3 | 0; 0.53; 0.0029; 0.47; 0; 0; 0; 0 | 0 1359 1 | 99352 180283 99375 | 99352 180283 99375 | 0 | 379010 |
| 5 | 0; 0.53; 0.0029; 0.47; 0; 0; 0; 0 | 0 1333 1 | 99273 179256 99307 | 99273 179256 99307 | 0 | 377836 |
| 5 | 0; 0.53; 0.0029; 0.47; 0; 0; 0; 0 | 0 1334 1 | 99274 179299 99306 | 99274 179299 99306 | 0 | 377879 |
| 8 | 0; 0.53; 0.0029; 0.47; 0; 0; 0; 0 | 0 1311 1 | 99161 177642 99177 | 99161 177642 99177 | 0 | 375980 |
| 8 | 0; 0.53; 0.0029; 0.47; 0; 0; 0; 0 | 0 1311 1 | 99158 177676 99191 | 99158 177676 99191 | 0 | 376025 |
| 15 | 0; 0.54; 0.0029; 0.46; 0; 0; 0; 0 | 0 1267 1 | 98909 174242 98922 | 98909 174242 98922 | 0 | 373335 |
| 15 | 0; 0.54; 0.0029; 0.46; 0; 0; 0; 0 | 0 1270 1 | 98911 174324 98940 | 98911 174324 98940 | 0 | 372175 |

*Feedback with Danger Zones:*

| Shift pause | Routing State Statistic | Switches per source [s1, s2, s3] | Packets sent by source [s1, s2, s3] | Packets at sink [s1, s2, s3] | Dropped Packets | total throughput |
|---|---|---|---|---|---|---|
| 1 | 0; 0.52; 0.0029; 0.48; 0; 0; 0; 0 | 0  1365  1 | 99424  181812  99451 | 99424 181812 99451 | 0 | 380687 |
| 1 | 0; 0.52; 0.0029; 0.48; 0; 0; 0; 0 | 0  1366  1 | 99424  181808  99451 | 99424 181808 99451 | 0 | 380683 |
| 2 | 0; 0.52; 0.0029; 0.47; 0; 0; 0; 0 | 0  1359  1 | 99388  180908  99415 | 99388 180908 99415 | 0 | 379711 |
| 2 | 0; 0.52; 0.0029; 0.47; 0; 0; 0; 0 | 0  1359  1 | 99388  180953  99415 | 99388 180953 99415 | 0 | 379756 |
| 3 | 0; 0.53; 0.0029; 0.47; 0; 0; 0; 0 | 0  1351  1 | 99351  180286  99370 | 99351 180286 99370 | 0 | 379007 |
| 3 | 0; 0.53; 0.0029; 0.47; 0; 0; 0; 0 | 0  1353  1 | 99344  180289  99378 | 99344 180289 99378 | 0 | 379011 |
| 5 | 0; 0.53; 0.0029; 0.47; 0; 0; 0; 0 | 0  1330  1 | 99272  179199  99302 | 99272 179199 99302 | 0 | 377773 |
| 5 | 0; 0.53; 0.0029; 0.47; 0; 0; 0; 0 | 0  1333  1 | 99278  179307  99305 | 99278 179307 99305 | 0 | 377890 |
| 8 | 0; 0.53; 0.0029; 0.47; 0; 0; 0; 0 | 0  1309  1 | 99156  177673  99192 | 99156 177673 99192 | 0 | 376021 |
| 8 | 0; 0.53; 0.0029; 0.47; 0; 0; 0; 0 | 0  1308  1 | 99159  177706  99171 | 99159 177706 99171 | 0 | 376036 |
| 15 | 0; 0.54; 0.0029; 0.46; 0; 0; 0; 0 | 0  1264  1 | 98911  173726  98938 | 98911 173726 98938 | 0 | 371575 |
| 15 | 0; 0.54; 0.0029; 0.46; 0; 0; 0; 0 | 0  1264  1 | 98902  175007  98928 | 98902 173742 98928 | 0,1265,0 | 372837, 371572 |

## Threshold:

*Estimate:*

| Threshold | Routing State Statistic | Switches per source [s1, s2, s3] | Packets sent by source [s1, s2, s3] | Packets at sink [s1, s2, s3] | Dropped Packets | total throughput |
|---|---|---|---|---|---|---|
| 3 | 0; 0.53; 0.0018; 0.47; 0; 0; 0; 0 | 0 2074 1 | 99350 180974 99369 | 99350 180974 99369 | 0 | 379693 |
| 3 | 0; 0.53; 0.0018; 0.47; 0; 0; 0; 0 | 0 2080 1] | 99348 180939 99368 | 99348 180939 99368 | 0 | 379655 |
| 3.5 | 0; 0.53; 0.0021; 0.47; 0; 0; 0; 0 | 0 1793 1 | 99347 180764 99369 | 99347 180764 99369 | 0 | 379480 |
| 3.5 | 0; 0.53; 0.0021; 0.47; 0; 0; 0; 0 | 0 1793 1 | 99341 180853 99364 | 99341 180853 99364 | 0 | 379558 |
| 4 | 0; 0.53; 0.0023; 0.47; 0; 0; 0; 0 | 0 1576 1 | 99344 180637 99368 | 99344 180637 99368 | 0 | 379349 |
| 4 | 0; 0.53; 0.0023; 0.47; 0; 0; 0; 0 | 0 1576 1 | 99341 180660 99364 | 99341 180660 99364 | 0 | 379365 |
| 4.5 | 0; 0.53; 0.0026; 0.47; 0; 0; 0; 0 | 0 1474 1 | 99351 180422 99376 | 99351 180422 99376 | 0 | 379149 |
| 4.5 | 0; 0.53; 0.0026; 0.47; 0; 0; 0; 0 | 0 1474 1 | 99349 180435 99378 | 99349 180435 99378 | 0 | 379162 |
| 5 | 0; 0.53; 0.0029; 0.47; 0; 0; 0; 0 | 0 1359 1 | 99352 180298 99376 | 99352 180298 99376 | 0 | 379026 |
| 5 | 0; 0.53; 0.0029; 0.47; 0; 0; 0; 0 | 0 1358 1 | 99349 180286 99379 | 99349 180286 99379 | 0 | 379014 |

*Feedback with Danger Zones:*

| Threshold | Routing State Statistic | Switches per source [s1, s2, s3] | Packets sent by source [s1, s2, s3] | Packets at sink [s1, s2, s3] | Dropped Packets | total throughput |
|---|---|---|---|---|---|---|
| 3 | 0; 0.53; 0.0018; | 0 2097 1 | 99350 180938 | 99350 180938 | 0 | 379657 |

| | 0.47; 0; 0; 0; 0 | | 99369 | 99369 | | |
|---|---|---|---|---|---|---|
| 3 | 0; 0.53; 0.0018; 0.47; 0; 0; 0; 0 | 0 2100 1 | 99349 180917 99368 | 99349 180917 99368 | 0 | 379634 |
| 3.5 | 0; 0.53; 0.0021; 0.47; 0; 0; 0; 0 | 0 1808 1 | 99350 180773 99370 | 99350 180773 99370 | 0 | 379493 |
| 3.5 | 0; 0.53; 0.0021; 0.47; 0; 0; 0; 0 | 0 1810 1 | 99350 180823 99370 | 99350 180823 99370 | 0 | 379543 |
| 4 | 0; 0.53; 0.0023; 0.47; 0; 0; 0; 0 | 0 1576 1 | 99350 180663 99370 | 99350 180663 99370 | 0 | 379383 |
| 4 | 0; 0.53; 0.0023; 0.47; 0; 0; 0; 0 | 0 1576 1 | 99352 180643 99365 | 99352 180643 99365 | 0 | 379360 |
| 4.5 | 0; 0.53; 0.0026; 0.47; 0; 0; 0; 0 | 0 1472 1 | 99344 180395 99376 | 99344 180395 99376 | 0 | 379115 |
| 4.5 | 0; 0.53; 0.0026; 0.47; 0; 0; 0; 0 | 0 1472 1 | 99352 180425 99377 | 99352 180425 99377 | 0 | 379154 |
| 5 | 0; 0.53; 0.0029; 0.47; 0; 0; 0; 0 | 0 1356 1 | 99345 180307 99369 | 99345 180307 99369 | 0 | 379021 |
| 5 | 0; 0.53; 0.0029; 0.47; 0; 0; 0; 0 | 0 1355 1 | 99349 180284 99376 | 99349 180284 99376 | 0 | 379009 |

## Allowed Drift:

*Feedback with Danger Zones:*

| Allowed Drift | Routing State Statistic | Switches per source [s1, s2, s3] | Packets sent by source [s1, s2, s3] | Packets at sink [s1, s2, s3] | Dropped Packets [s1, s2, s3] |
|---|---|---|---|---|---|
| 0.3 | 0; 0.53; 0.0029; 0.47; 0; 0; 0; 0 | 0 1349 1 | 99351 180645 99376 | 99351 180645 99376 | 379372 |
| 0.3 | 0; 0.53; 0.0029; 0.47; 0; 0; 0; 0 | 0 1343 1 | 99350 180680 99379 | 99350 180444 99379 | 0, 236, 0 |
| 0.03 | 0; 0.53; 0.0029; 0.47; 0; 0; 0; 0 | 0 1360 1 | 99340 180300 99374 | 99340 180300 99374 | 0 |
| 0.03 | 0; 0.53; 0.0029; 0.47; 0; 0; 0; 0 | 0 1355 1 | 99351 180285 99379 | 99351 180285 99379 | 0 |
| 0.003 | 0; 0.53; 0.0029; 0.47; 0; 0; 0; 0 | 0 1353 1 | 99350 180287 99378 | 99350 180287 99378 | 0 |
| 0.003 | 0; 0.53; 0.0029; 0.47; 0; 0; 0; 0 | 0 1353 1 | 99349 180276 99377 | 99349 180276 99377 | 0 |